\documentclass[conference]{IEEEtran}
\IEEEoverridecommandlockouts

\usepackage{xcolor}
\usepackage{booktabs} %\toprule \midrule 
\usepackage{subcaption}

\usepackage{cite}
\usepackage{amsmath,amssymb,amsfonts}
\usepackage{algorithmic}
\usepackage{graphicx}
\usepackage{textcomp}
\usepackage{xcolor}
\usepackage[english]{babel}
\def\BibTeX{{\rm B\kern-.05em{\sc i\kern-.025em b}\kern-.08em
    T\kern-.1667em\lower.7ex\hbox{E}\kern-.125emX}}
\definecolor{NVD}{HTML}{2CA02C}
\definecolor{AMZ}{HTML}{FF7F0E}
\definecolor{IBM}{HTML}{1F77B4}
\definecolor{RTX}{HTML}{D62728}
\usepackage{soul}

\begin{document}
%
%
% The "title" command has an optional parameter, allowing the author to define a "short title" to be used in page headers.
\title{Performance Analysis of Deep Learning Workloads on Leading-edge Systems}

% \affiliation{%
%     \institution{Brookhaven National Laboratory}
%   \streetaddress{P.O. Box 5000}
%   \city{Upton}
%   \state{New York}
%   \postcode{11793}
% }

\author{\IEEEauthorblockN{Yihui Ren}
%\IEEEauthorblockN{2\textsuperscript{nd} Shinjae Yoo}
%\IEEEauthorblockN{3\textsuperscript{rd} Adolfy Hoisie}
\IEEEauthorblockA{\textit{Computational Science Initiative}\\
\textit{Brookhaven National Laboratory}\\
yren@bnl.gov}
\and
\IEEEauthorblockN{Shinjae Yoo}
\IEEEauthorblockA{\textit{Computational Science Initiative}\\
\textit{Brookhaven National Laboratory}\\
sjyoo@bnl.gov}
\and
\IEEEauthorblockN{Adolfy Hoisie}
\IEEEauthorblockA{\textit{Computational Science Initiative}\\
\textit{Brookhaven National Laboratory}\\
ahoisie@bnl.gov}
}

\maketitle

\begin{abstract} 

This work examines the performance of leading-edge systems designed for machine learning computing, including the NVIDIA DGX-2, Amazon Web Services (AWS) P3, IBM Power System Accelerated Compute Server AC922, and a consumer-grade Exxact TensorEX TS4 GPU server. Representative deep learning workloads from the fields of computer vision and natural language processing are the focus of the analysis. Performance analysis is performed along with a number of important dimensions. Performance of the communication interconnects and large and high-throughput deep learning models are considered. Different potential use models for the systems as standalone and in the cloud also are examined. The effect of various optimization of the deep learning models and system configurations is included in the analysis. 

\end{abstract}

%
% A "teaser" image appears between the author and affiliation information and the body 
% of the document, and typically spans the page. 
%   \Description{Enjoying the baseball game from the third-base seats. Ichiro Suzuki preparing to bat.}

%
% This command processes the author and affiliation and title information and builds
% the first part of the formatted document.
% \maketitle

\section{Introduction}

%% Here are some new citations. 

%% \ray{
%% \begin{itemize}
%%     \item Adofly's paper \cite{tallent_evaluating_2017}
%%     \item OepnACC CUDA \cite{herdman_accelerating_2012}
%%     \item reference workloads for modern deep learning \cite{adolf_fathom:_2016}
%%     \item Asynchronous multi-gpu programming model \cite{ben-nun_groute:_2017}
%%     \item performance analysis and optmization ong GPUs using CUDA \cite{habich_performance_2011}
%%     \item knights landing intel xeon phi \cite{sodani_knights_2016}
%%     \item performance on multi-gpu with expether \cite{nomura_performance_2014}
%%     \item scaling deep learning workloads dgx-1p and intel knights \cite{gawande_scaling_2017}
%%     \item ultra-performance pascal gpu nvlink \cite{foley_ultra-performance_2017}
%%     \item parallel performance \cite{malony_parallel_2011}
%% \end{itemize}
%% }
The growth of machine learning and deep learning (DL) extends across all data
analytical application areas, impacting many disciplines and markets. Hence,
their practical use potential appears exponential and seemingly unbounded. In
turn, the ever-insatiable need for computing resources for these workloads has
led to the development of computer architectures and systems designed to
improve machine learning
performance~\cite{sodani_knights_2016,adolf_fathom:_2016,ciregan_multi-column_2012,graves_speech_2013,malony_parallel_2011,nvidia_nvidia_2017-1,tallent_evaluating_2017,li_tartan:_2018}.
As the presently preferred architectures for machine learning application
workloads, GPU-based systems are an important exemplar in this category.

%While research analyzing GPU performance has some history~\cite{malony_parallel_2011,habich_performance_2011}, such GPU analyses have been more recent for DL workloads~\cite{tallent_evaluating_2017,adolf_fathom:_2016}. 
This work evaluates the performance of two important types of DL algorithms on four leading-edge GPU-based systems. Specifically, we consider convolutional neural network (CNN) algorithms, such as AlexNet and ResNet, mostly used in computer vision and attention-mechanism-based algorithms for natural language processing on the NVIDIA DGX-1 and DGX-2, IBM Power System AC922, and Exxact TensorEX TS4. Moreover, we analyze a cloud-based Amazon Web Services (AWS) P3dn use mode for the DGX-1 and compare DL performance against standalone use for the other systems considered. 

GPU-based systems are especially well suited for DL workloads as proven in
practice and in scientific
publications~\cite{raina_large-scale_2009,chetlur_cudnn:_2014,ciregan_multi-column_2012}.
Briefly, this stems from their single-instruction multiple-data (SIMD) nature
and arithmetic intensity of the algorithms mapping well to available floating
point operations (FLOPS) on GPUs; availability of large amounts of
high-bandwidth memory that allows for data access at fast rates and low
latency; and to high-speed interconnects that afford communication at high
bandwidth with minimal contention. The first three examples of leading-edge
systems considered herein use the NVIDIA Tesla V100 GPU with different
topologies of the NVLink interconnect. The Exxact TS4 is configured with the
consumer-grade GeForce RTX 2080 Ti GPU, which is popular among AI researchers,
developers, and hobbyists. Section~\ref{subsec:hardware} describes the systems
and their key architectural characteristics in more detail.

Section~\ref{sec:dlm} details how DL models considered are trained, the fundamental arithmetic operations involved during training, and their effects on different hardware systems. Specifically, Section~\ref{subsec:cv} dissects CNN models for computer vision, while Section~\ref{subsec:nlp} explores the state-of-the-art Bidirectional Encoder Representations from Transformers (BERT) model for natural language processing (NLP)~\cite{devlin_bert:_2018}.

%{\ray Adolfy: Add here one paragraph as to the importance of computer vision and NLP. In the
%description we need to say something about computation and communication
%requirements of these specific algorithms. A detailed description of the
%algorithms and benchmarks is included in Section ??.
%}

%{\cg Ray Added:

The detailed performance analysis is done along a few important dimensions. Section~\ref{subsec:comm} presents the performance of key global communication kernels used in the benchmarks considered. Section~\ref{subsec:largeDLM} discusses performance and scalability of large and high-throughput DL models. Section~\ref{subsec:cdp} compares performance when the benchmarks are expressed in an easy-to-code multi-GPU architecture enabled by system software described in Section~\ref{subsec:soft}.

\section{Environment}
\subsection{Hardware Environment}\label{subsec:hardware}
As part of this work, the following systems were put to the test: NVIDIA DGX-1V and DGX-2 (\textcolor{NVD}{DGX-2}), IBM Power System AC922 (\textcolor{IBM}{IBM-P9}), AWS P3dn (\textcolor{AMZ}{AWS P3}), and Exxact TensorEX TS4 (\textcolor{RTX}{RTX}). Henceforth, the systems will be referenced using their respective abbreviations noted in parentheses. For added convenience, a consistent color scheme and geometric shape are maintained for each system represented in figures throughout this work (green diamond, DGX-2; blue square, IBM-P9; orange triangle, AWS P3; red circle, RTX). Of note, the AWS P3 essentially is a DGX-1V as shown in the communication bandwidth test depicted in Section~\ref{subsec:comm}.

Before delving into the details of each system, we first introduce the key architectural component: the NVIDIA Tesla V100 GPU. 

\subsubsection*{Tesla V100}
The Tesla V100 GPU~\cite{nvidia_nvidia_2017} is a building block for three of the four systems under consideration. The V100 GPU has 640 Tensor cores and 5,120 CUDA cores with 32 GB (or 16 GB) HBM2 GPU memory (900 GB/s bandwidth). It can achieve 15.7 TFLOPS for single-precision performance. For direct inter-device (GPU-to-GPU) communication, the V100 has six NVLink-2.0 fabric supporting 25 GB/s per link, per data direction. Therefore, each V100 has the ability to communicate with other GPU devices at 150 GB/s unidirectional (or 300 GB/s bidirectional) bandwidth. The high bandwidth of inter-node communication is crucial for training deep neural network models across multiple devices. 

\begin{figure}[ht]
    \begin{subfigure}[t]{1.0\columnwidth}
        \centering
        \includegraphics[width=1.0\columnwidth]{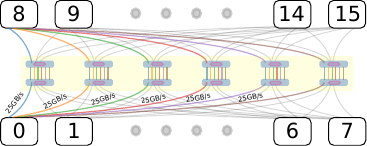}
        \caption{DGX-2 NVSwitch Crossbar}\label{fig:dgx2} 
        \vspace*{5mm}
    \end{subfigure}
    \begin{subfigure}[t]{1.0\columnwidth}
        \centering
        \includegraphics[width=1.0\columnwidth]{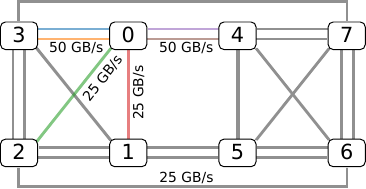}
        \caption{DGX-1V and AWS P3 Hybrid Cube-Mesh Topology}\label{fig:dgx1} 
    \vspace*{5mm}
    \end{subfigure}
    \begin{subfigure}[t]{1.0\columnwidth}
        \centering
        \includegraphics[width=1.0\columnwidth]{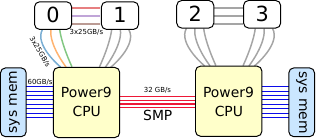}
        \caption{IBM AC922 Model 8335-GTH NVLink-enabled POWER9 CPU}\label{fig:ac922} 
    \end{subfigure}
    \caption{\label{fig:comm}GPU-to-GPU Communication Topology.
    Each Tesla V100 GPU has six NVLink ports with unidirectional communication bandwidth of 25 GB/s per port. Numerically labeled boxes represent different GPU devices. The six NVLinks from device-0 are colored differently.}
\end{figure}

\subsubsection*{\textcolor{NVD}{DGX-2}}
The bulk of the DGX-2's computation capacity is from 16 V100 (32 GB) GPUs evenly distributed on two baseboards and connected via 12 on-node switches, or \textit{NVSwitch}~\cite{nvidia_nvidia_2018}. Each NVSwitch has 18 NVLink ports (16 in use) and supports 900 GB/s bidirectional peak bandwidth. Eight NVLink ports are connected to different GPU devices (one per link) on the same baseboard, whereas the other eight NVLink ports are connected to the matching NVSwith ports on the other baseboard (Figure~\ref{fig:dgx2}). This network
connectivity affords communications at a bandwidth of up to 150 GB/s per direction. Any two V100 GPUs can establish full bandwidth (up to 150 GB/s per direction) communication using all six NVLink ports. The specific DGX-2 tested in this work has two hyper-threaded 24-core Intel Xeon 8168 CPUs (96 logic cores in total) with base frequency of 2.7 GHz, 1.5 TB system memory, and 30 TB NVMe SSD in eight-way RAID0.

\subsubsection*{\textcolor{AMZ}{AWS P3}}
AWS' P3dn.24xlarge instance is similar to the NVIDIA DGX-1V system~\cite{nvidia_nvidia_2017-1} and is equipped with eight Tesla V100 (32 GB) GPUs connected in a hybrid cube-mesh topology (Figure~\ref{fig:dgx1}). The hybrid cube-mesh topology leads to each node having four immediate neighbors. This is a legacy design following the previous DGX-1P system, where the Tesla P100 GPU featured only four NVLink ports. Two of the four neighbors are connected to two links each, while the other two connect to one only. 
%This leads to an uneven peer-to-peer communication pattern: for two device connected via one
%single NVLink such as between device-0 and device-1, the unidirectional bandwidth is only upto
%25GB/s; whereas it is 50GB/s between device-0 and device-4.
To connect two P3 systems, AWS provides network connection bandwidth up to 100 Gbits/s. The caveat is that this limit can be reached only for multi-flow connections. The single-flow bandwidth is 10 Gbits/s (1.25 GB/s). 
%% At the time of writing this article, AWS has launched preview the Elastic Fabric
%% Adapter\footnote{https://aws.amazon.com/about-aws/whats-new/2018/11/introducing-elastic-fabric-adapter/}
%% which supports better inter-node communication. 
The specific AWS P3 systems tested in this effort have two hyper-threaded 24-core Intel Xeon 8175M CPUs (96 logic cores in total) with base frequency of 2.5 GHz, 768 GB system memory, and 2 TB ephemeral NVMe SSD. Section~\ref{subsec:comm} shows that the NVIDIA DGX-1V system is analogous to the AWS P3. Thus, we include only the results for the AWS P3.

\subsubsection*{\textcolor{IBM}{IBM-P9}}
% The IBM Power System AC922~\cite{caldeira_ibm_2018} (Model 8335-GTH) server tested is equipped with four Tesla V100 (16 GB) GPUs (Figure~\ref{fig:ac922}). The tested AC922 server has two IBM POWER9 hyper-threaded 20-core CPUs (160 logic cores in total) with base frequency of 2.4 GHz. IBM's POWER9 CPU is NVLink-enabled. Each CPU has six direct NVLink connections to GPUs (three per GPU), enabling a 75 GB/s unidirectional communication bandwidth to each GPU. In addition, there are three NVLink fabrics connecting two GPUs directly. If the GPUs are not connected to the same CPU, communications must route through the inter-CPU symmetric multiprocessing (SMP) cable with unidirectional bandwidth of 32 GB/s. The POWER9 CPU connects to the system main memory with accumulated (eight channels) unidirectional bandwidth of 60 GB/s. 
The IBM Power System AC922~\cite{caldeira_ibm_2018} (Model 8335-GTH) server tested is equipped with four Tesla V100 (32 GB) GPUs (Figure~\ref{fig:ac922}). The tested AC922 server has two IBM POWER9 hyper-threaded 20-core CPUs (160 logic cores in total) with base frequency of 2.3 GHz and max frequency of 3.8 GHz. IBM's POWER9 CPU is NVLink-enabled. Each CPU has six direct NVLink connections to GPUs (three per GPU), enabling a 75 GB/s unidirectional communication bandwidth to each GPU\@. In addition, there are three NVLink fabrics connecting two GPUs directly. If the GPUs are not connected to the same CPU, communications must route through the inter-CPU symmetric multiprocessing (SMP) cable with unidirectional bandwidth of 32 GB/s. The POWER9 CPU connects to the system main memory with accumulated (eight channels) unidirectional bandwidth of 60 GB/s. The tested system has four nodes, connected via high-bandwidth (24~GB/s unidirectional) InfiniBand. All of the nodes use IBM General Parallel File System (GPFS) with block size of 1 MB and bandwidth of approximately 18~GB/s.

%%The specific AC922 server tested in this work uses a network file system and may present some difficulties under intensive data loading workloads. Removed by AH.

\subsubsection*{\textcolor{RTX}{RTX}}
The Exxact TensorEX 4U server (TS4-1598415-DPN) is equipped with eight NVIDIA consumer-grade GeForce RTX 2080 Ti GPUs~\cite{nvidia_nvidia_2018-1}. Each RTX
2080 Ti GPU has 4352 CUDA cores and 11 GB GDDR6 GPU memory with 616 GB/s memory bandwidth. It can reach a peak performance of 13.4 TFLOPS for single-precision performance, or about 85.4\% of the V100 GPU's peak performance. The specific server tested in this work has two hyper-threaded 12-core Intel Xeon 4116 CPUs (48 logic cores in total) with base frequency of 2.1 GHz. All eight GPUs are connected via a PCIe bus. Compared to other high-end V100 GPU-based solutions, the RTX GPU cards are a unique feature for this system. As such, we refer to this system as {\textcolor{RTX}{RTX}}.
%https://blog.exxactcorp.com/exploring-the-complexities-of-pcie-connectivity-and-peer-to-peer-communication/

\subsection{Software Environment}\label{subsec:soft}
Because of its popularity among AI researchers, its well-designed user interface, and native support for NVIDIA communication and computation backend kernels and MPI, we use the PyTorch DL platform. To maintain a consistent and reproducible software environment, we use docker containers, which also alleviate the difficulty in migrating the DL models to other hardware systems and reduce the performance differences introduced by distinct software environments. For the x86 architecture (Intel Xeon CPU) systems, including DGX-1, DGX-2, AWS P3, and RTX, we use the NVIDIA official PyTorch docker image (NVCR)\footnote{nvcr.io/nvidia/pytorch:18.11-py3} as the base software environment. For the ppc64le architecture (IBM POWER9 CPU) system, IBM-P9, we use the PowerAI v1.6~\cite{furmanek_powerai_2019}.

Nevertheless, to ensure our work is reproducible, Table~\ref{tab:env} lists the exact library versions of the NVIDIA docker and the PowerAI v1.6. The NVIDIA CUDA library is a
programming interface to NVIDIA GPUs for parallel computing, while NVIDIA's cuDNN (deep neural network) library provides device-level optimized, neural-network-related backend kernels. The NVIDIA NCCL (collective communication) library provides a multi-GPU communication interface, supporting several communication means, such as NVLink, PCIe, and Ethernet.

\begin{table}[th]
  \caption{Software Environment}
  \label{tab:env}
  \centering
  \begin{tabular}{lrr}
    \toprule
    Library               &   NVIDIA NVCR       & IBM PowerAI          \\
    \midrule
    {\verb|PyTorch|}      & {\verb|1.0.0a0 |}   & {\verb|1.1.0   |}    \\
    {\verb|CUDA   |}      & {\verb|10.0.130|}   & {\verb|10.1.168|}    \\
    {\verb|cuDNN  |}      & {\verb|7.401   |}   & {\verb|7.501   |}    \\
    {\verb|NCCL   |}      & {\verb|2.307   |}   & {\verb|2.407   |}    \\
  \bottomrule
\end{tabular}
\end{table}

\section{Deep Learning Models}\label{sec:dlm}

\subsection{Data Movement and Communication Between Devices}

Deep learning is a data-driven modeling approach. The training process, known as \textit{stochastic gradient descent}, consists of numerous iterations of feeding data to the model and adjusting the model parameters to reduce the predefined loss. At each iteration, a batch of data is selected at random (without replacement). The data are loaded from the hard drive to the host memory, and, sometimes, preprocessing data-augmentation procedures are applied using CPU threads, such as randomly flipping images or adjusting image sizes. Then, the preprocessed batch is sent to the GPU memory via PCIe bus.

The bulk of actual computation usually is done on one or multiple GPUs. In the multiple GPU case, the execution is done in a SIMD fashion so each GPU has an exact replica of the neural network model and applies the exact executions on different sampled data batches. In the ideal case, the throughput would grow linearly with the number of GPUs. At the end of every iteration, all of the model replicas require synchronization. This synchronization is done by a collective communication using NCCL. Most of the results in this work use the NCCL all-reduce kernel. Therefore, the two major factors affecting the time cost of communication are: 1) the inter-device communication bandwidth and 2) number of model parameters.

For this work, we have selected several representative DL models to cover
different ranges of parameters, computation-communication ratios, application
domains, and various types of neural network DL layers. Because of the vast
number of potential DL models, we are unable to test all of them exhaustively.
However, by providing detailed descriptions and computation characteristics for
these select models, readers should be able to easily estimate the performance
(in terms of computation efficiency not model accuracy) of other models as the
fundamental types of numeric operations are comparable. As computer vision and
NLP are the two most successful application domains for DL, we choose the
AlexNet model and ResNet model from the computer vision domain and BERT model
from NLP to represent examples of DL methods in these areas. We analyze the
models in terms of their number of trainable parameters and operations. The
former affects the memory footprint as well as the inter-device communication
costs, while the latter impacts the on-device computation time. The computation
cost per iteration scales linearly with the number of instances per sampled
data batch, known as the \textit{batch size}.  However, the actual computation
cost depends on many other factors.  Table~\ref{tab:dpm} provides a summary of
the number of parameters and operations per instance
for all of the models presented in this work.

\begin{table}
    \caption{Tested Deep Learning Models}
    \label{tab:dpm}
    \centering
  \begin{tabular}{lrr}
    \toprule
    Model Name &         Param.  &         Ops/ins. \\ % & ratio $\Gamma$\\
    \midrule
    AlexNet    & 61.10 M             & 0.72  G      \\ %     & 11.78 \\
    ResNet18   & 11.69 M             & 1.83  G      \\ %     & 156.54\\
    ResNet50   & 25.56 M             & 4.14  G      \\ %     & 161.97\\
    ResNet101  & 44.55 M             & 7.88  G      \\ %     & 176.88\\
    ResNet152  & 60.19 M             & 11.62 G      \\ %     & 193.06\\
    BERT-SWAG  & 109.5 M             & 0.19  G      \\ %     & 1.74  \\
    BERT-SQuAD & 109.5 M             & 2.87  G      \\ %     & 26.21 \\
  \bottomrule
\end{tabular}
\end{table}

\subsection{Computer Vision}\label{subsec:cv}
The goal of computer vision is to make computers gain high-level ``understanding'' of images. To evaluate if a program (AI model) truly ``understands'' the image, researchers have developed different evaluation tasks to measure its comprehension. One type of these tasks, known as \textit{image classification}, provides an image to the program and asks about which predefined class the image belongs to. For example, the MNIST (handwritten digit database) asks the program to tell it which digit, from 0 to 9, the grayscale image (28-by-28 pixels) belongs to. This is considered one of the simplest computer vision tasks, and traditional machine learning methods, such as the support vector method, have reached 99.2\% accuracy~\cite{lecun_gradient-based_1998}. The ImageNet Large Scale Visual Recognition Challenge, or ILSVRC~\cite{russakovsky_imagenet_2015}, a much more challenging image classification test, was introduced in 2010. It contains 1000 predefined classes (including 60 different dog breeds) and more than a million training images. The best-performing model in the first ILSVRC (2011) achieved only about a 25\% top-five error rate.\footnote{\textit{Top-five error rate}. For each test image, the algorithm is allowed to give five predictions. If any of the five predictions match to the ground truth, it is considered a hit.} In 2012, AlexNet~\cite{krizhevsky_imagenet_2012}, considered the first modern CNN-based model, successfully reduced the top-five error rate to 16.4\%. In 2015, ResNet~\cite{he_deep_2016} further reduced the error rate to 3.57\%. It also introduced residual blocks to mitigate the ``vanishing gradient problem'' when the neural network becomes too deep.

A deep neural network is a stack of multiple neural network layers, usually varying kinds. Each layer takes the previous layer's output as its input, where both input and output are tensors. A \textit{Linear} layer is one of the simplest kind, a matrix of size $c_i \times c_o$, where $c_i$ and $c_o$ are the number of input and output channels. Therefore, the number of parameters of a Linear layer is on the order of $O(c_i c_o)$ or $c_o(c_i+1)$ to be precise where ``1'' is the bias term. The operation performed by a Linear layer essentially is a general matrix-matrix multiplication (GEMM). In most cases, the multiplier matrix (input) has a dimension of $B \times c_i$, and the multiplicand matrix (Linear layer weights) has a dimension of $c_i \times c_o$. As such, the number of operations for a batch size $B$ is $B \times (c_i+1) \times c_o$. One could deduce that the operation-to-parameter ratio $\Gamma$ for a Linear layer is $B$: $\Gamma_\text{Linear} = B$, implying that computation cost grows linearly with the number of parameters in the Linear layer and batch size.

A two-dimensional convolutional (Conv2D) layer consists of $c_o$ kernels of size $c_i\times k\times k$. Therefore, the exact number of parameters of a Conv2D layer is $c_o (k^2 c_i+1)$. A kernel is simply a small tensor applied to the input tensor in a sliding-window fashion, where the step size is called \textit{the stride}. When the stride is greater than one, the input tensor is downsampled in the spatial dimension. The number of operations for a Conv2D layer can be calculated by considering the number of times the kernel has been applied and the cost of applying each kernel. Applying a Conv2D kernel on the input tensor of size $B \times c_i \times H_i \times W_i$ is meant to perform a tensor dot product of $c_i \times k^2$ on every pixel of the spatial dimension $H\times W$.

For simplicity, assume the striding step is 1, and padding is $\lfloor k/2 \rfloor$ such that the spatial dimension is unchanged $H_o = H_i$ and $W_o = W_i$.  Thus, each kernel has been applied $H_o \times W_o$ times.\footnote{ Note that by setting striding greater than one, fewer kernel operations will be applied, which can reduce the spatial dimension (downsampling). Whereas, by setting the space between kernel points (dilation), the spatial dimension (upsampling) can increase. The computation cost analysis is similar.}
For each kernel application at every pixel level, a GEMM operation is performed, which costs $C \equiv c_o(c_i k^2+1)$. Therefore, in total, the number of operations of the Conv2D layer is $H_o\times W_o \times C$. Because the number of parameters of a Conv2D layer is also $C$, the operation-to-parameter ratio $\Gamma$ for Conv2D layer is $\Gamma_\text{Conv2D} = BH_oW_o$. As in the case of the Linear layer, the total number of operations scales with the batch size. Yet, in contrast to the Linear layer, the total number of operations also depends on the spatial dimension of the output tensor. Each parameter of a Conv2D layer has been operated $H_oW_o$ more times than a parameter in a Linear layer.

AlexNet consists of five Conv2D layers of $\sim 2^{21}$ parameters in total, two hidden Linear layers ($\sim 2^{25}$), and one output Linear layer ($\sim 2^{22}$). The Linear layer also uses an order of magnitude more parameters. Compared to AlexNet, ResNet consists almost entirely of Conv2D layers, except the final Linear layer for classification output.
%The kernel dimension of Conv2D are also relatively small: $1\times1$ or $3\times3$.
The sub-types of ResNet models are labeled as ResNet{\bf X}, where {\bf X} represents the total number of parameterized layers (Conv2D and Linear). The choices of X in the original paper~\cite{he_deep_2016} are 18, 34, 50, 101, and 152. ResNet18 serves as a high-throughput (small number of operations), low-accuracy model because of the small amount of parameters, while ResNet152 has the highest accuracy but slowest training throughput. Using ResNet50 for ImageNet data (1000-way classification) as a concrete example, the model contains about $2^{24.6}$ parameters, where only $2^{21}$ are from the Linear layer. As discussed, each parameter of a Conv2D layer contributes a factor of $H_o \times
W_o$ more operations than one in a Linear layer. As such, ResNet has a much higher operation-to-parameter ratio than AlexNet.

\subsection{Natural Language Processing}\label{subsec:nlp}
NLP is another successful application of DL techniques. Some NLP tasks include speech recognition, translation, speech-to-text (and vice versa), and question-and-answer systems. In the pre-DL era, NLP was dominated by hidden Markov models~\cite{gales_application_2007}. Mikolov et al.~\cite{mikolov_distributed_2013} introduced a DNN-based word embedding model to represent words as vectors based on their context. Namely, similar words would have comparable context around them and end up closer in the vector space. This approach provides a meaningful way to represent non-numeric entities, i.e., words, as numeric vectors and provides a foundation for solving a diverse range of NLP tasks. Graves et al.~\cite{graves_speech_2013} developed a deep recurrent-neural-network-based approach to perform automatic speech recognition and broke the TIMIT phoneme recognition benchmark record~\cite{garofolo_darpa_1993}. By the end of 2016, all major technology companies had adopted the DNN-based approach for their speech recognition systems. Vaswani et al.~\cite{vaswani_attention_2017} introduced the attention mechanism into NLP tasks and demonstrates its superior performance in natural language translation tasks.

The particular NLP model in this work, BERT, uses bidirectional transformers ~\cite{devlin_bert:_2018} and exceeded 11 NLP benchmark records in November 2018.\footnote{As of March 2019, OpenAI and Microsoft have released their model challengers to BERT.}

The BERT model has two training phases: 1) \emph{pre-training} and 2) \emph{fine-tuning}. In the pre-training phase, BERT uses the semi-supervised sequence learning approach~\cite{dai_semi-supervised_2015} by masking out a random word in a sentence. Unlike other previous unidirectional approaches, BERT tries to predict the masked word from both directions. Training is done on large unlabeled corpora, such as the English Wikipedia (2,500 million words). Herein, this pre-trained model is known as the \textit{base-model}. In the task-specific fine-tuning phase, the base-model connects with a classification Linear layer designed for the specific task. The data used for fine-tuning are labeled and much smaller compared to the large corpora~\cite{radford_improving_2018}.
%The BERT base-model consists of $L$ Transformer blocks with $H$ hidden layers and $A$ self-attention heads.
%Each Transformer block consists of {\ray TODO: decide if we need the detailed explanation}.
The majority of attention mechanism operations are matrix multiplication and layer-wise normalization. For details regarding how the attention mechanism works, readers can refer to several available guides.\footnote{http://nlp.seas.harvard.edu/2018/04/03/attention.html.}$^{,}$\footnote{https://jalammar.github.io/illustrated-transformer/.}

We use the pre-trained BERT base-model and fine-tune it for two specific NLP tasks: SWAG and Stanford Question Answering Dataset (SQuAD). The SWAG~\cite{zellers_swag:_2018} is a multi-choice task. Given a situation described by a sentence as \textit{input}, the model is asked to select the most plausible scenario that happens next among multiple choices. The SQuAD~\cite{rajpurkar_squad:_2016} is a Question Answering task, where a pair that includes a question and a relevant paragraph (containing the answer) is provided and the model is tasked to find the answer in the given paragraph.

Although the base model is the same, to fully cover the training data, different max-seq-length is used. We use max-seq-length of 80 for SWAG and 384 for SQuAD. As the max-seq-length determines the attention span, it takes more operations to perform the SQuAD task. Table~\ref{tab:dpm} features the number of model parameters and estimated operations of BERT-SWAG and BERT-SQuAD, respectively. Of note, our benchmark code is modified from the source code.\footnote{https://github.com/huggingface/pytorch-pretrained-BERT}

\section{Performance Analysis}

This section details the performance analysis of DL workloads using the four systems (already described) under consideration. The all-important communication performance is first presented. Given the different workload characteristics, the analysis is done separately for large-scale and high-throughput models. Performance details for an increasingly popular code expression (due to ease of coding)---PyTorch's \textit{On-node Data Parallel}~\cite{paszke_automatic_2017}---also is included. 

\subsection{Communication Performance}\label{subsec:comm}
As shown in Section~\ref{subsec:hardware}, leading-edge systems implement
various direct high-bandwidth inter-device communication topologies based on
NVLink. The
NCCL\footnote{https://developer.nvidia.com/nccl}
provides MPI-like primitives for multi-GPU and multi-node collective
communications. The library is optimized for NVIDIA GPU devices to achieve high
communication bandwidth over NVLink and PCIe (when necessary). NCCL supports
collective communication primitives, such as all-reduce, all-gather,
reduce-scatter, reduce, and broadcast.

As the most relevant communication kernels occurring in the benchmarks
considered, all-reduce and broadcast are examined for performance using
NVIDIA's NCCL-tests code.\footnote{https://github.com/NVIDIA/nccl-tests/release/tag/v1.0.0}
Results are presented normalized to the "bus bandwidth," a concept described by
NVIDIA in the NCCL-tests.\footnote{Described in detail
here:\\https://github.com/NVIDIA/nccl-tests/blob/master/doc/PERFORMANCE.md.}
Bus bandwidth is obtained by applying a normalization divider of the measured
bandwidth\\($\text{``message size''} / \text{time}$) different for each
communication kernel to reflect its communication complexity and topological
mapping to the network. Because the bus bandwidth reflects how optimally the
hardware is used, it provides a consistent and normalized way to compare the
results with the theoretical peak bandwidth, including across different
communication primitives.

\begin{figure}[th]
    \begin{subfigure}[th]{1.0\columnwidth}
        \centering
        \includegraphics[width=1.0\columnwidth]{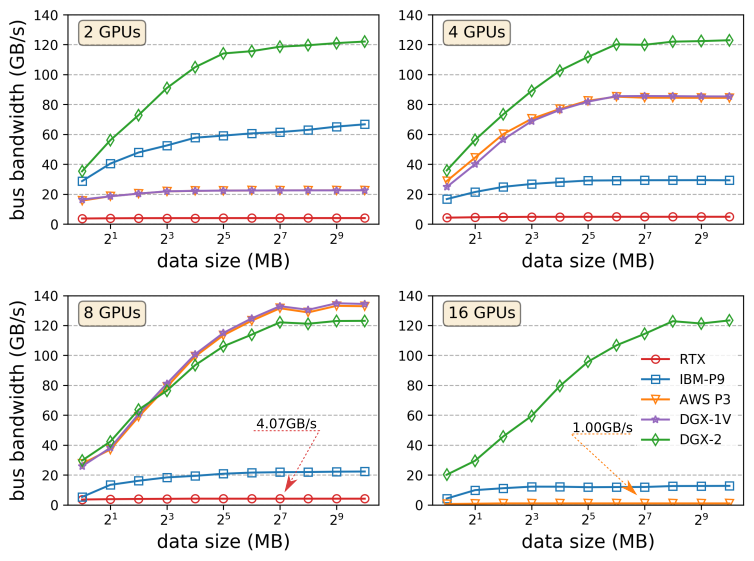}
        \caption{\label{fig:allreduce} All-reduce}
        \vspace*{5mm}
    \end{subfigure}
    \begin{subfigure}[th]{1.0\columnwidth}
        \centering
        \includegraphics[width=1.0\columnwidth]{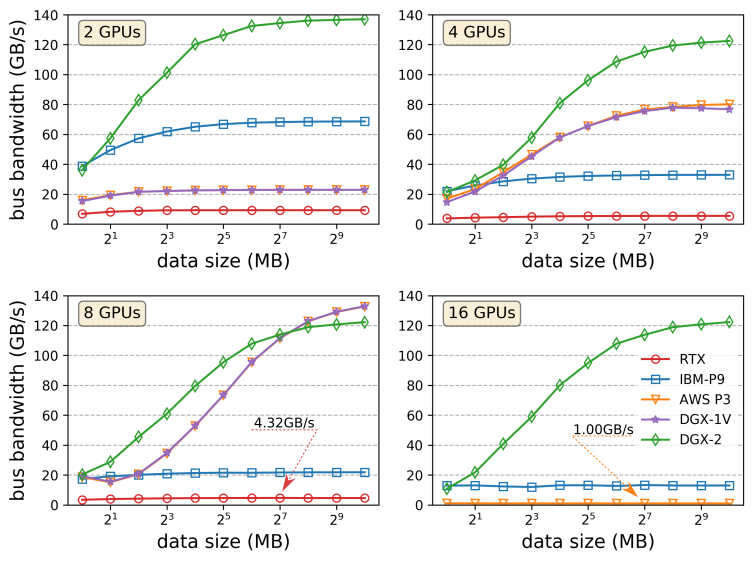}
        \caption{\label{fig:broadcast} Broadcast}
    \end{subfigure}
  \caption{\label{fig:nccl} Communication Bus Bandwidth.}
\end{figure}

In this work, data size varies from 1 MB to 1 GB, which covers the communication needs for synchronizing model parameters. Each data point is averaged over 500 iterations, except for the case of 16 GPUs using two AWS P3s, which is averaged over 50 iterations due to the slow inter-node Ethernet connection. Figure~\ref{fig:nccl} illustrates the results.
%The all-reduce NCCL kernel and broadcast  is primarily used in PyTorch distributed data-parallel
%training scenario, whereas the reduce and broadcast are used in PyTorch DataParallel module wrapper.
%At the end of every iteration, all model replicas need to
%synchronize by accumulating the gradients from other devices for each
%parameters of the model, thus an all-reduce communication with the ``sum'' operation. 

The DGX-2 consistently achieves 120~GB/s for large message sizes, regardless of the number of GPUs involved in the communications. This can be attributed to the NVSwitch's link bandwidth and contention properties (described in Section~\ref{subsec:hardware}).

% \sout{
% DGX-2 can achieve consistent 120GB/s for any number of the GPU devices. 
% As a reminder, DGX-2 connect all 16 GPUs with the cluster of 12 NVSwitches, Figure~\ref{fig:dgx2}.
% It does not matter how many GPUs are involved in the communication nor the type of collective communication.
% However, in most tested cases, the peak bandwidth of 150GB/s is not reached.
% The closest it can get is about 140GB/s when broadcasting messages from one GPU to the other (2GPUs), Figure~\ref{fig:broadcast}.
% }

The AWS P3 and DGX-1V yield analogous, if not exactly the duplicate, results
because they share the same hybrid cube-mesh topology (refer to
Figure~\ref{fig:dgx1}). Because of the heterogeneity of this topology, the
measured peak bandwidth depends on the devices involved in the communication.
In the case of two GPUs, the test employs device-0 and device-1, which are
connected via a single NVLink that offers 25 GB/s theoretical unidirectional
bandwidth. For four GPUs, device-0 to -3 are used, and the NVLinks
connecting to device-4 to -7 are not. The observed bandwidth is about
80 GB/s. For eight GPUs, the DGX-1 surpasses the DGX-2 in the all-reduce tests
(Figure~\ref{fig:allreduce}). In the broadcast test
(Figure~\ref{fig:broadcast}), the crossover occurs when the message size
exceeds 256~MB\@. While these results may seem unexpected due to the higher
bandwidth and topological richness of the NVSwitch compared to the NVLink, the
actual explanation stems from the communication protocol changes introduced on
the NVSwitch~\cite{tierney_nccl_2019}. Here, posted requests are converted to non-posted, which, in
turn, requires \textit{acks} at the expense of bandwidth in the reverse
direction. This is not the case on the DGX-1V without NVSwitch. With access to
only one DGX-1, the 16 GPU case was done on AWS P3. The two AWS-P3dn nodes are
connected via a 100 Gbits/s multi-flow Ethernet connection. The experimental
setup in the AWS cloud allowed for only a single flow (review
Section~\ref{subsec:hardware}) with a peak bandwidth of 1.25 GB/s. In this
case, the communication bandwidth clearly is bottlenecked by the slow Ethernet
connection.

IBM-P9 uses half of the NVLinks for CPU-GPU communication (Figure~\ref{fig:ac922}). This leaves three NVLinks to connect device-0 and device-1. In the case of two GPUs, the measured bus bandwidth of 70 GB/s is quite close to the theoretical peak of 75 GB/s. However, with four GPUs, the bus bandwidth reduces to about 30 GB/s, matching the theoretical SMP bus bandwidth of 32 GB/s when connecting two POWER9 CPUs. Higher count GPU configurations on the IBM P9 (eight- and 16-GPU) exhibit lower bus bandwidth (Figure~\ref{fig:nccl}). This achieved performance is due to NCCL not being optimized for the InfiniBand interconnect.

The RTX system does not use NVLink technology, and all eight RTX 2080Ti GPUs connect through a PCIe bus. Therefore, the communication bandwidth is throttled down by the PCIe bus. Despite its inferior communication performance, the RTX system serves as the baseline for other systems.

\subsection{Performance of Deep Learning Workloads  }\label{subsec:largeDLM}
Computation performance is measured in terms of the model training throughput:
the average number of training samples, or \textit{instances}, the system can
process per second. For each different combination of models, batch sizes, and
number of GPUs, time intervals are measured between consecutive iterations
during training. For computer vision DL models, each model runs for 200
iterations. For the BERT models, the reported throughput is averaged over one
training epoch.\footnote{One epoch is defined as going through the entire data
set once.} The initial iterations are excluded from the statistics due to
memory allocation overhead. All of the models in this secion are
represented in single precision (FP32).

Distributed data-parallel training with asynchronous data prefetching is used. Each GPU is associated with $j$ data-fetching CPU processes using CUDA streams. In these tests $j=4$. This allows data to be loaded and preprocessed asynchronously and concurrently on the CPUs while the GPUs are in use. Every GPU device holds a replica of the model and applies the model on different data batches. At each iteration's conclusion, all GPUs synchronize their parameter gradients via an all-reduce NCCL operation. Then, all model replicas individually update their parameters using the gradients. The computer vision models are trained on the ILSVRC ImageNet data set, while BERT models are fine-tuned on task-specific data sets, SWAG and SQuAD (introduced in Section~\ref{subsec:nlp}).

As the system performance characteristics vary for different models, we group models such as ResNet101(152) and BERT as \textit{large} DL models and those with high throughput, e.g., ResNet18(50) and AlexNet, as \textit{high-throughput} DL models. The large DL model results are discussed in this sub-section, while high-throughput DL models are addressed in Section~\ref{subsec:smallDLM}. For added clarity, the bar plots featured in this section depict systems ordered from left to right, corresponding to the system order posed in legends (inset from top to bottom). 

\subsubsection{\textbf{Performance Analysis of Large Deep Learning Models}}\label{subsubsec:scaleLDM}

\begin{figure}[ht]
    \centering
    \begin{subfigure}[t]{0.49\columnwidth}
        \includegraphics[width=1\columnwidth]{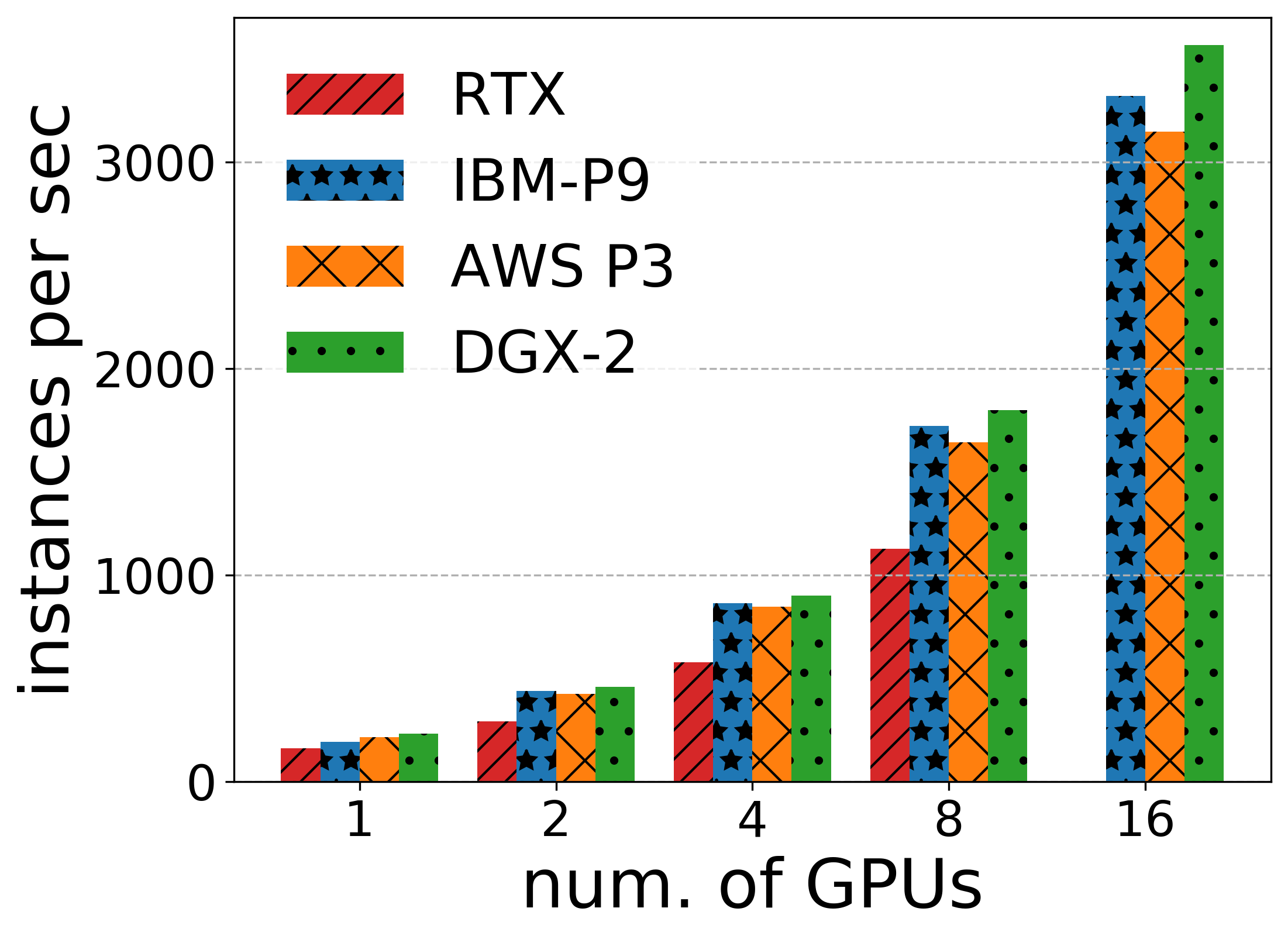}
        \caption{\label{fig:scale-resnet101}ResNet101}
    \end{subfigure}
    \begin{subfigure}[t]{0.49\columnwidth}
        \includegraphics[width=1\columnwidth]{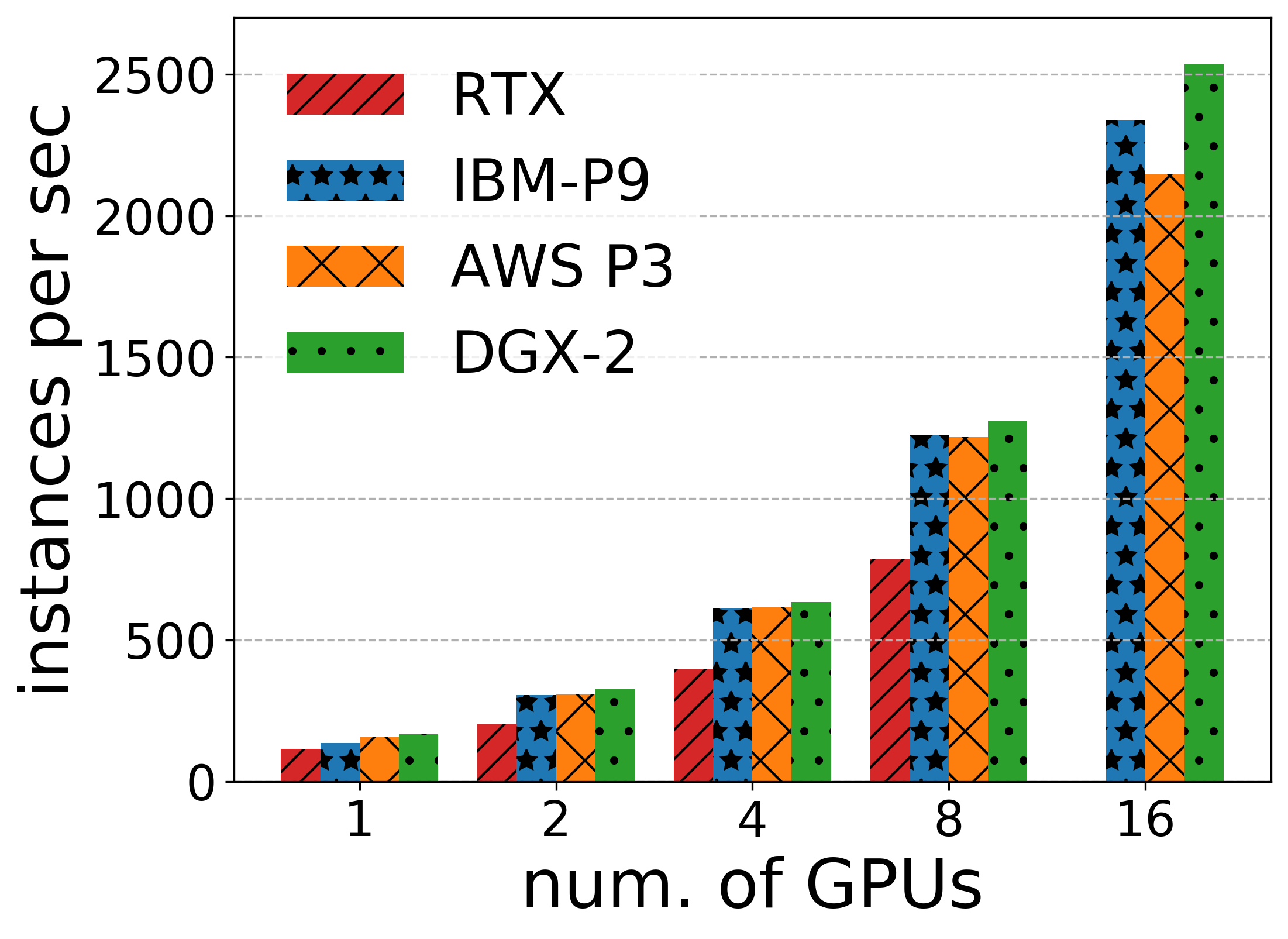}
        \caption{\label{fig:scale-resnet152}ResNet152}
    \vspace*{5mm}
    \end{subfigure}
    \begin{subfigure}[t]{0.49\columnwidth}
        \includegraphics[width=1\columnwidth]{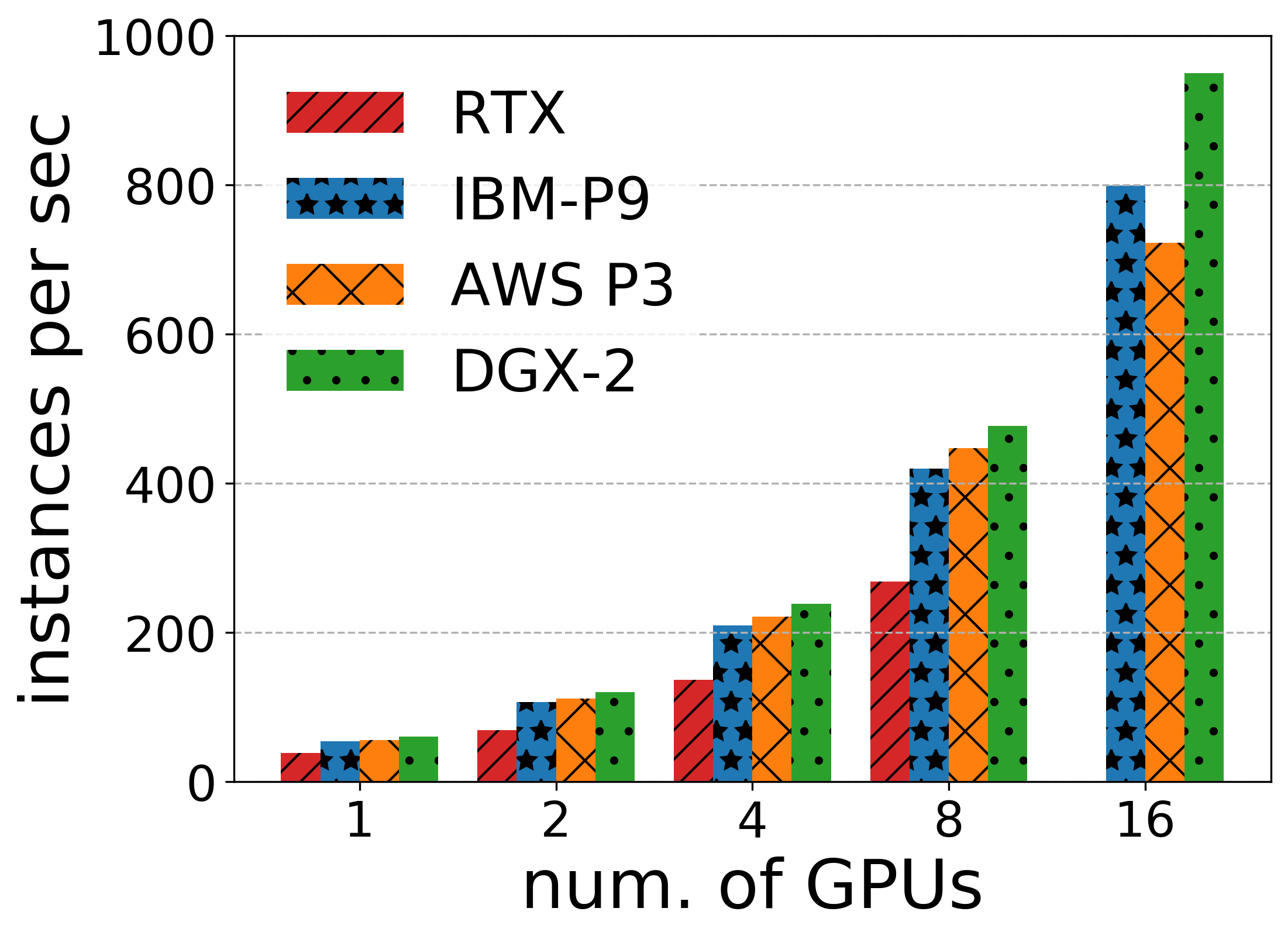}
        \caption{\label{fig:scale-bertswag}BERT-SWAG}
    \end{subfigure}
    \begin{subfigure}[t]{0.49\columnwidth}
        \includegraphics[width=1\columnwidth]{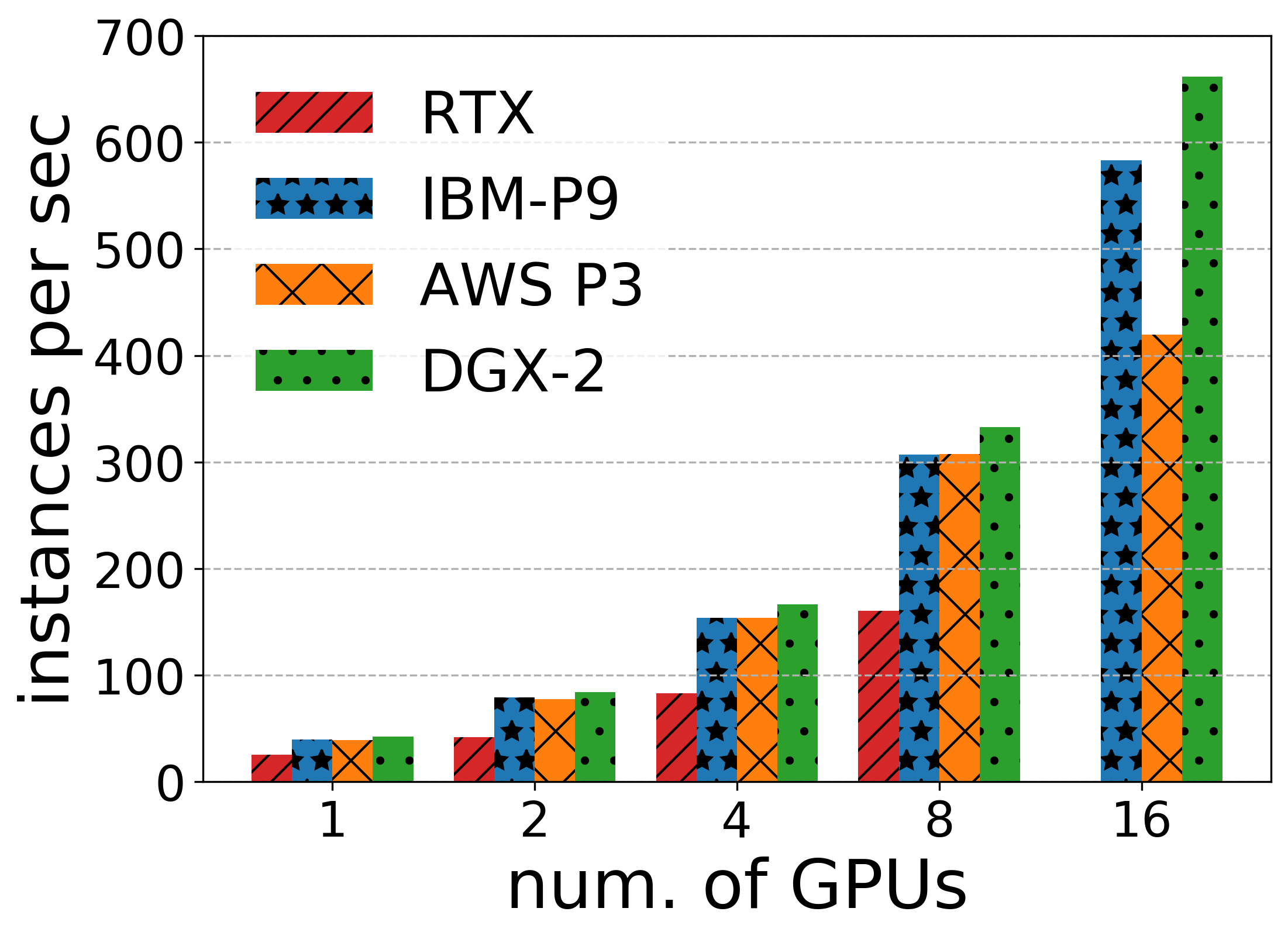}
        \caption{\label{fig:scale-bertsquad}BERT-SQuAD}
    \end{subfigure}
    \caption{\label{fig:scale-LDM}
        Training Throughput of Large DL Models on RTX, IBM-P9, AWS P3, and DGX-2.
      }
\end{figure}

%Performance including scalability of large-scale DL workloads is discussed in this section.
Initially, the absolute throughput values of large DL models, e.g., ResNet101, ResNet152, BERT-SWAG, and BERT-SQuAD, are examined
(Figure~\ref{fig:scale-LDM}). As the amount of communication for synchronization depends on the number of model parameters and not on the batch size, we choose the largest batch size that can fit into the 32 GB of memory of a single V100 GPU to achieve the best possible scaling results. Specifically, the batch sizes used are: 128 per GPU for ResNet101 and ResNet152, 64 for BERT-SWAG, and 32 for BERT-SQuAD\@. 

Across all four systems, the DGX-2 and AWS P3 have similar performance up to
eight GPUs. This is expected as both systems have the same V100 GPUs and are
connected via high-bandwidth (over 120 GB/s) NVLinks. However, when 16 GPUs are
in use, two AWS P3s communicate through a relatively slow Ethernet connection
(about 1 GB/s measured).
Figures~\ref{fig:scale-bertswag}~and~\ref{fig:scale-bertsquad} reveal the
differences in performance, especially in BERT models where the number of
parameters is large.
Given its high-bandwidth inter-node communication network, 
the IBM P9 exhibits similar performance to DGX-2 all the
way to up to a 16 GPU configuration. 

The RTX server has 11 GB of DDR6 GPU memory. Hence, the batch sizes are even smaller: one-quarter of the size when using 32 GBs on the V100 GPU on all other systems. Specifically, the batch size for ResNet101 and ResNet152 is 64, BERT-SQuAD is 8, and BERT-SWAG is 16. This leads to a quadrupling of the amount of communication for the same total of computed instances. RTX's slow inter-device communication via a PCIe bus further exacerbates its performance degradation. For example, in the case of 1 GPU, RTX can reach about 65.82\% throughput of the DGX-2 averaged over four DL models, yet merely 57.27\% in the case of eight GPUs (see Table~\ref{tab:rtx-vs-dgx2}). Hence, the RTX server is the least efficient system for large model distributed training.

\begin{figure}[ht]
    \centering
    \begin{subfigure}[t]{0.49\columnwidth}
        \includegraphics[width=1.0\columnwidth]{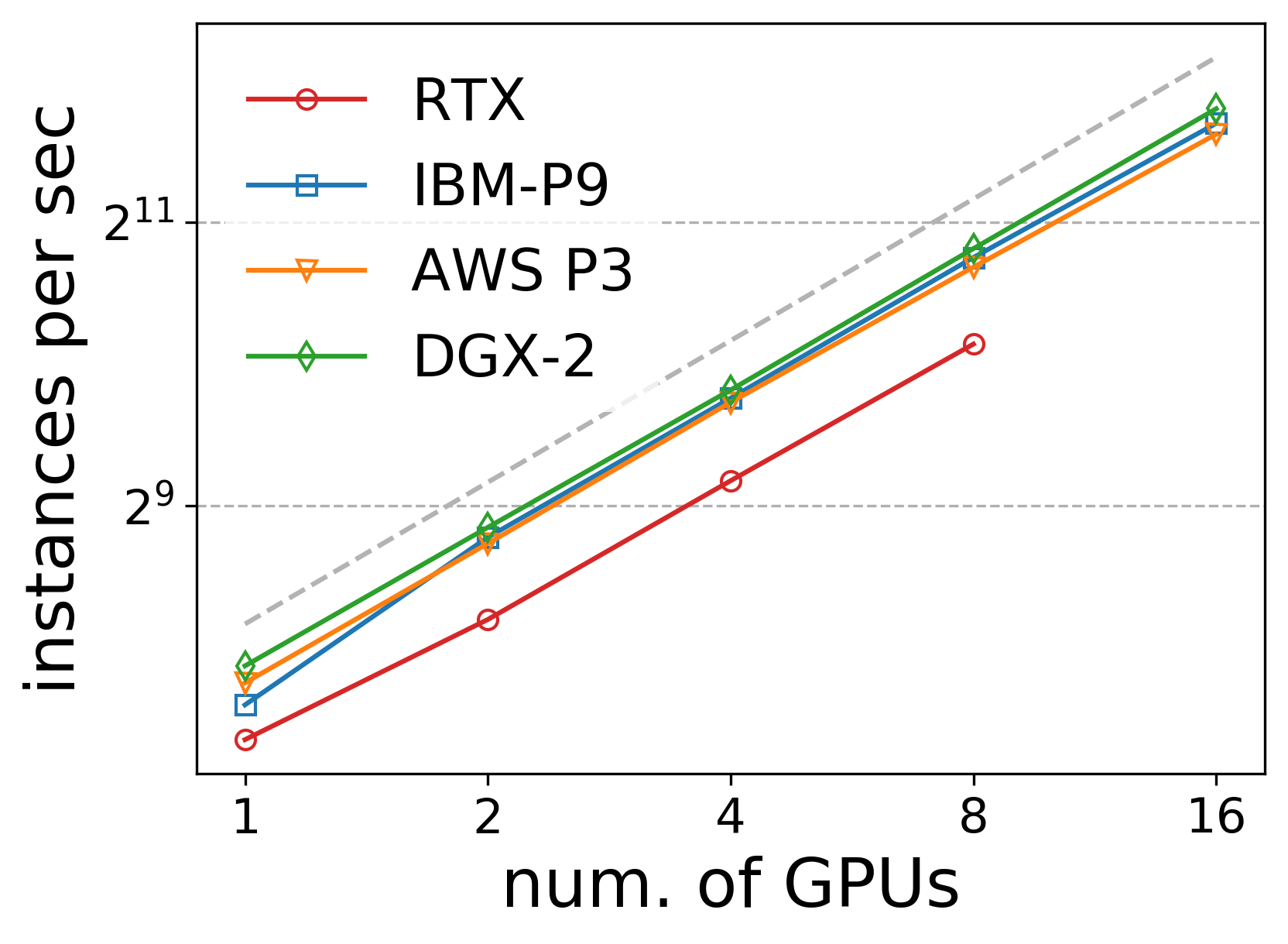}
        \caption{\label{fig:loglogresnet101} ResNet101}
    \end{subfigure}
    \begin{subfigure}[t]{0.49\columnwidth}
        \includegraphics[width=1.0\columnwidth]{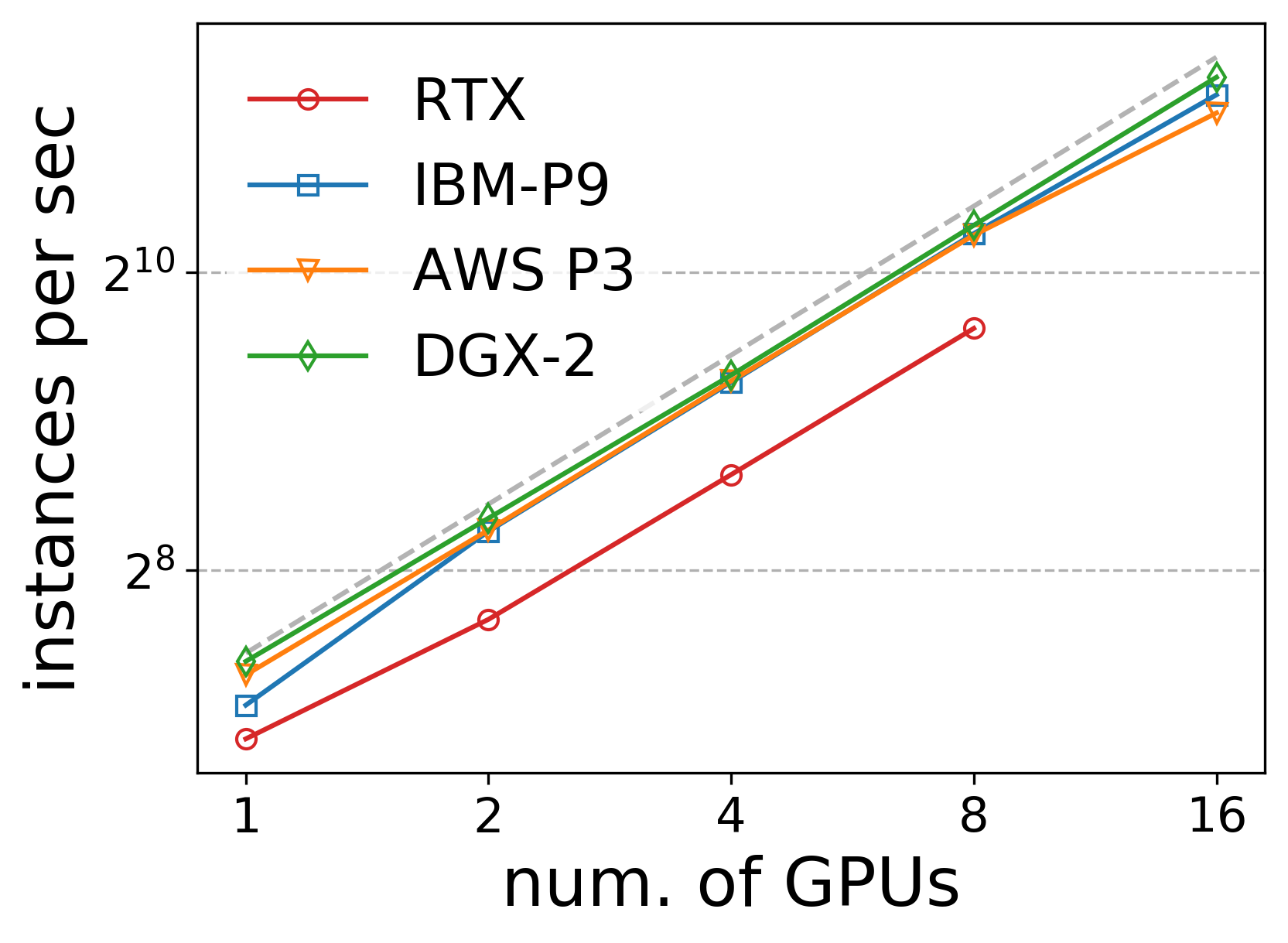}
        \caption{\label{fig:loglogresnet152} ResNet152}
    \vspace*{5mm}
    \end{subfigure}
    \begin{subfigure}[t]{0.49\columnwidth}
        \includegraphics[width=1.0\columnwidth]{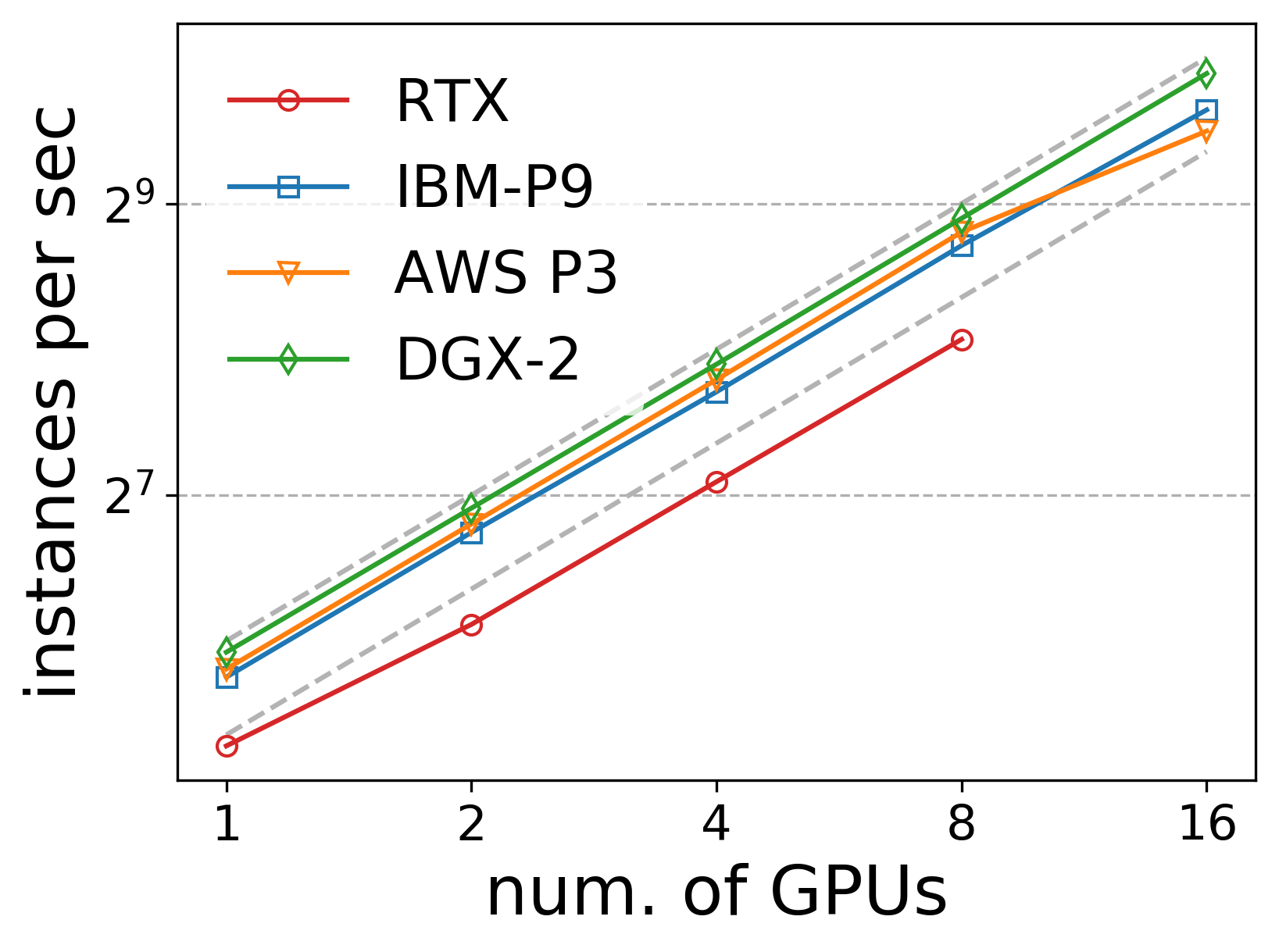}
        \caption{\label{fig:loglogbertswag} BERT-SWAG}
    \end{subfigure}
    \begin{subfigure}[t]{0.49\columnwidth}
        \includegraphics[width=1.0\columnwidth]{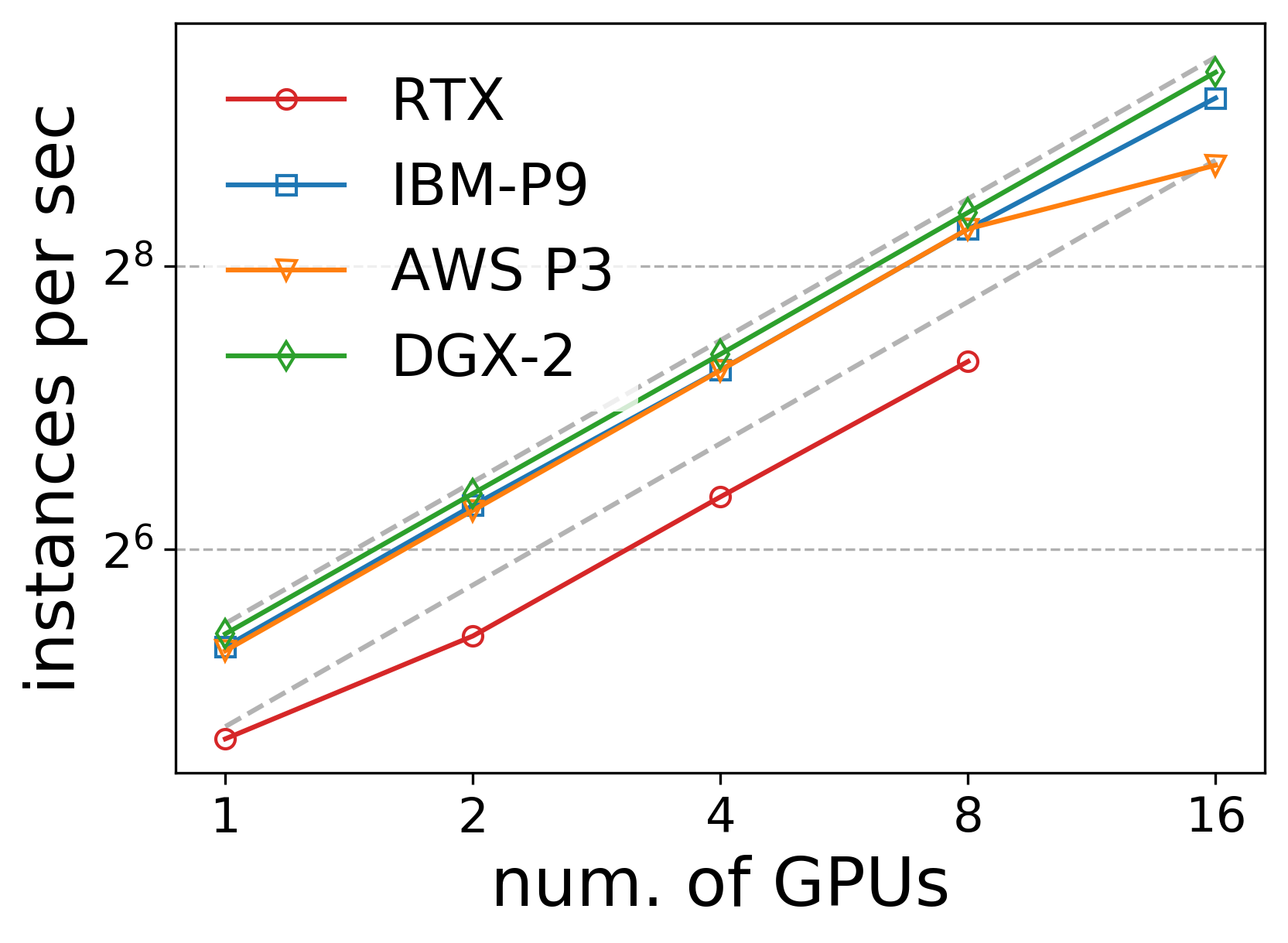}
        \caption{\label{fig:loglogbertsquad} BERT-SQuAD}
    \end{subfigure}
    \caption{\label{fig:loglog-LDM} Linear Scaling in Log-Log Scale.
        Gray dashed lines are linear scaling reference lines.
    }
\end{figure}

To examine the scaling more closely throughout the full span of GPU configurations, we plot the throughput for all DL models in a log-log scale (Figure~\ref{fig:loglog-LDM}), where the dashed reference line depicts linear scalability. If the measured throughput follows the reference line, or maintains a constant gap, it has good parallel scalability. The DGX-2 exhibits good scalability on all four models, whereas AWS P3 shows linear scalability up to eight GPUs. For the RTX, there is a significant drop from one GPU to two GPUs in terms of scalability because one GPU computation does not require model synchronization, while that cost does apply for multiple GPU configurations.
%\adf{WHY\@? ASSUME some cache effect?
%But beyond 2 GPUs, it appears to scale linearly. }
%\ray{1 GPU does not require model synchronization. But 2 GPU needs that. RTX uses slow PCIe.}
%\cg{TODO: add something about IBM P9 once we have results}

\begin{table}[hb]
    \caption{\label{tab:rtx-vs-dgx2} Instances per second for RTX relative to DGX-2}
    \centering
    \begin{tabular}{lrrrr}
        \toprule
        Model Name & 1 GPU   & 2 GPUs  & 4 GPUs  & 8 GPUs  \\
        \midrule
        AlexNet      & 78.19\% & 63.01\% & 53.41\% & 47.95\% \\
        ResNet18     & 73.50\% & 69.13\% & 64.39\% & 54.80\% \\
        ResNet50     & 67.97\% & 62.67\% & 62.97\% & 61.75\% \\
        Average      & 73.22\% & 64.94\% & 60.26\% & 54.83\%\\
        \midrule
        ResNet101    & 69.70\% & 63.72\% & 64.15\% & 62.69\% \\
        ResNet152    & 69.73\% & 62.45\% & 62.96\% & 61.90\% \\
        BERT-SWAG    & 64.04\% & 57.52\% & 57.20\% & 56.25\% \\
        BERT-SQuAD   & 59.81\% & 49.79\% & 49.74\% & 48.22\% \\
        Average      & 65.82\% & 58.37\% & 58.51\% & 57.27\% \\
        \midrule
        Overall avg. & 68.99\% & 61.19\% & 59.26\% & 56.22\% \\
        \bottomrule
    \end{tabular}
\end{table}

% \subsection{Performance Analysis of High-throughput Deep Learning Models}\label{subsec:smallDLM}

\begin{figure*}[ht]
    \centering
    \begin{subfigure}[t]{0.49\columnwidth}
        \includegraphics[width=1\columnwidth]{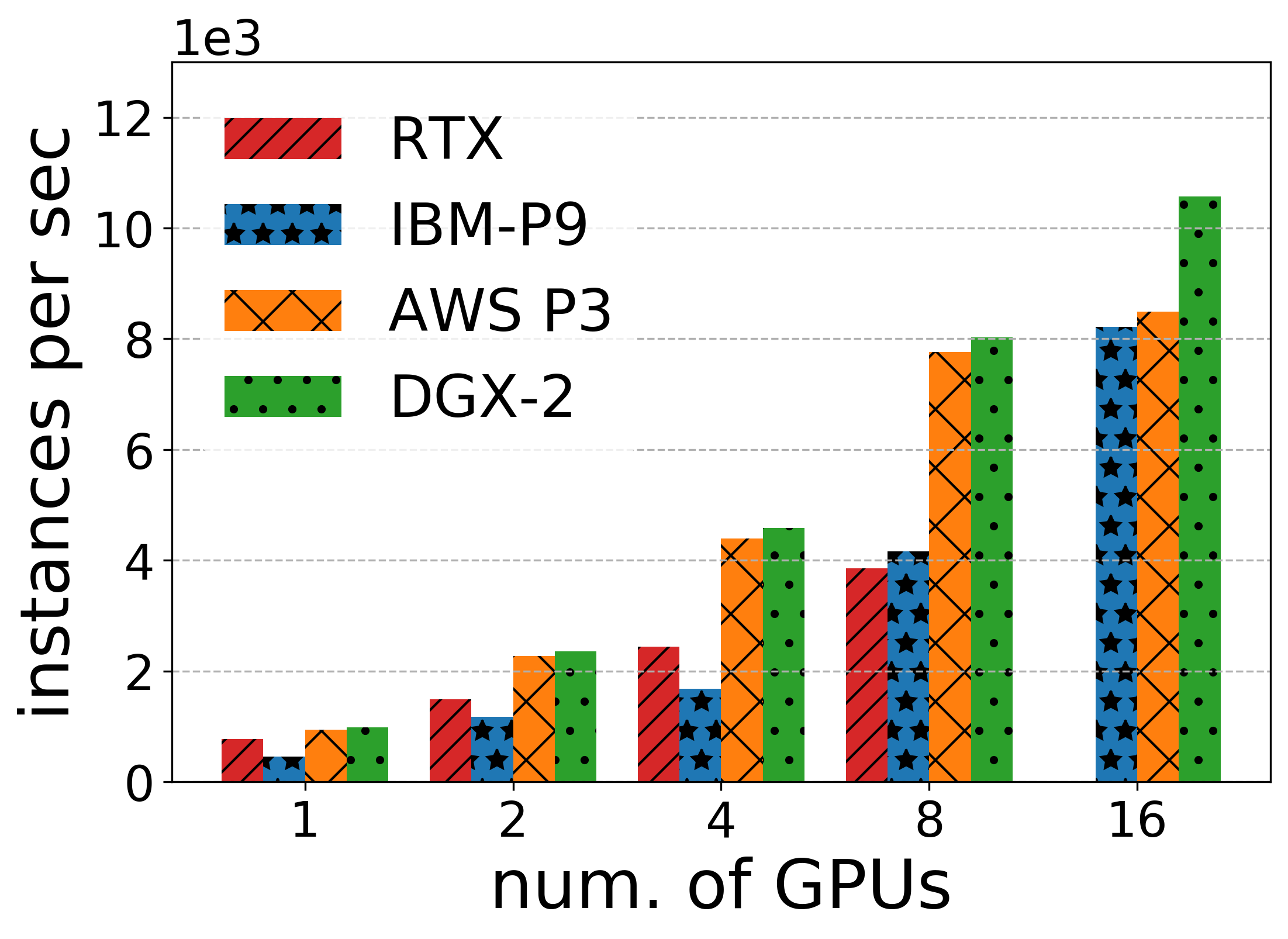}
        \caption{\label{fig:scale-alexnet} AlexNet}
    \end{subfigure}
    \begin{subfigure}[t]{0.49\columnwidth}
        \includegraphics[width=1\columnwidth]{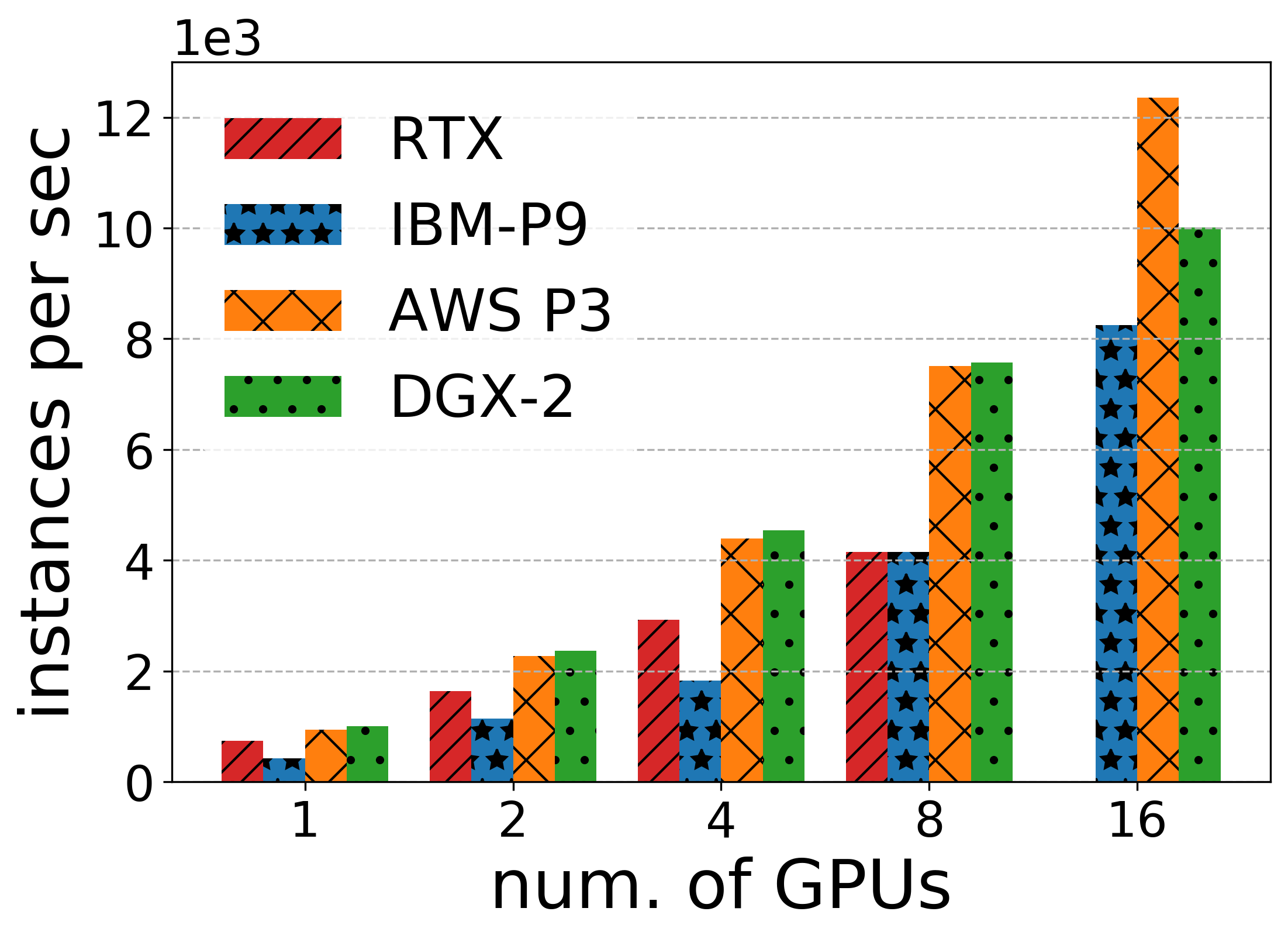}
        \caption{\label{fig:scale-resnet18}ResNet18}
    %\vspace*{5mm}
    \end{subfigure}
    \begin{subfigure}[t]{0.49\columnwidth}
        \includegraphics[width=1\columnwidth]{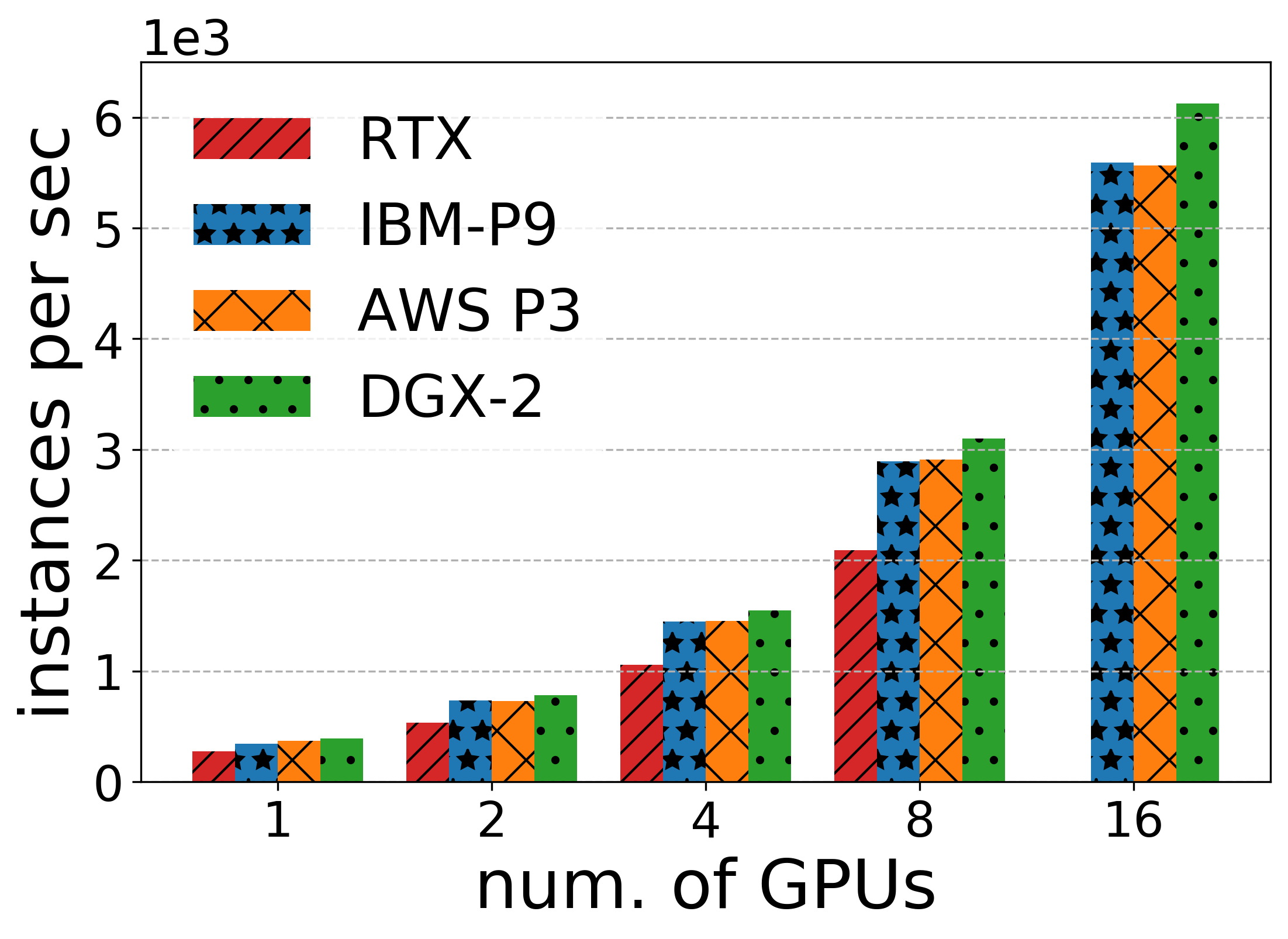}
        \caption{\label{fig:scale-resnet50}ResNet50}
    \end{subfigure}
    \caption{\label{fig:scale-SDM}
        Training Throughput of High-throughput DL Models on RTX, IBM-P9, AWS P3, and DGX-2.
  }
\end{figure*}

\begin{figure*}[ht]
    \centering
    \begin{subfigure}[t]{0.49\columnwidth}
        \includegraphics[width=1\columnwidth]{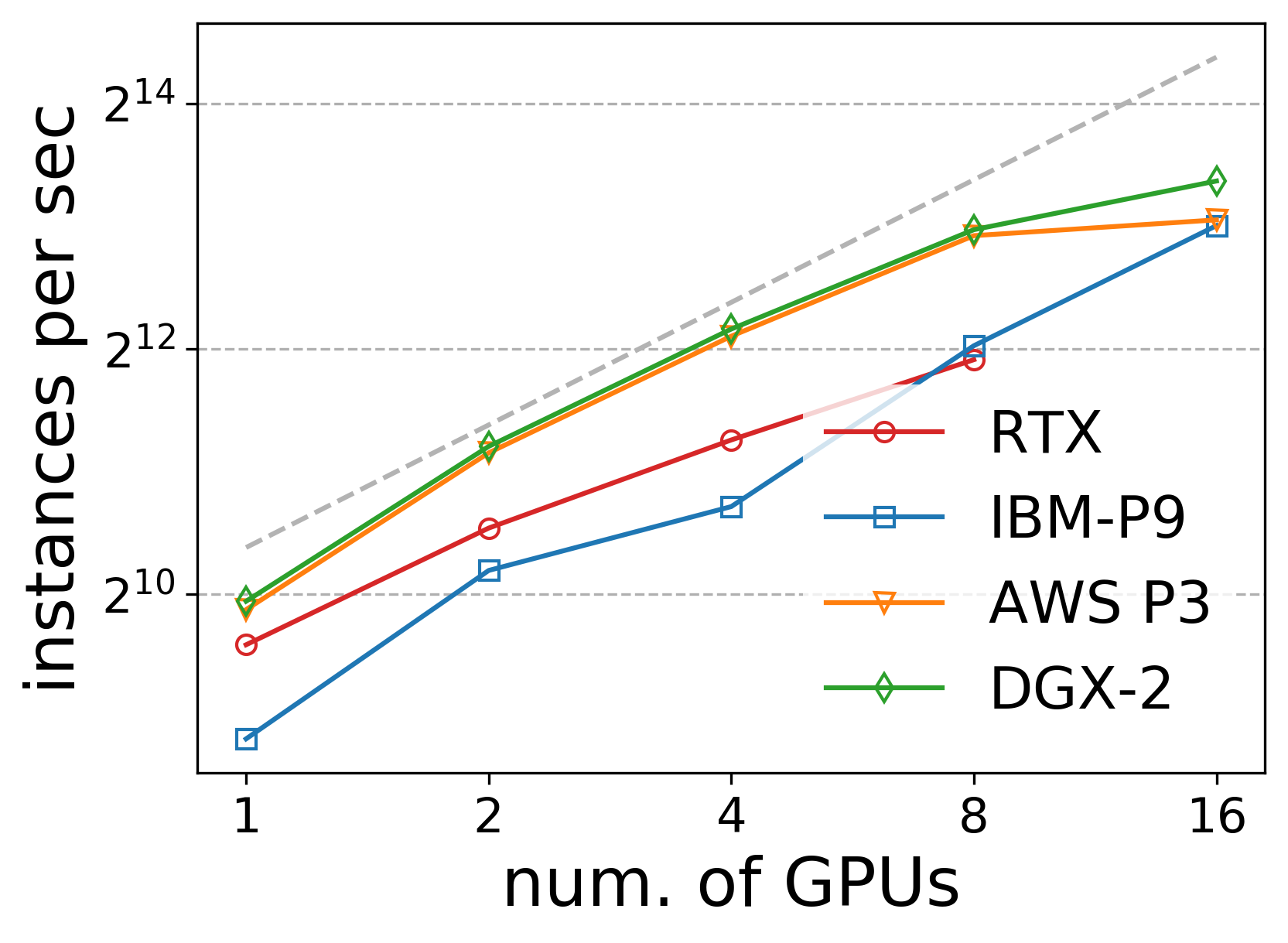}
        \caption{\label{fig:loglog-alexnet} AlexNet}
    \end{subfigure}
    \begin{subfigure}[t]{0.49\columnwidth}
        \includegraphics[width=1\columnwidth]{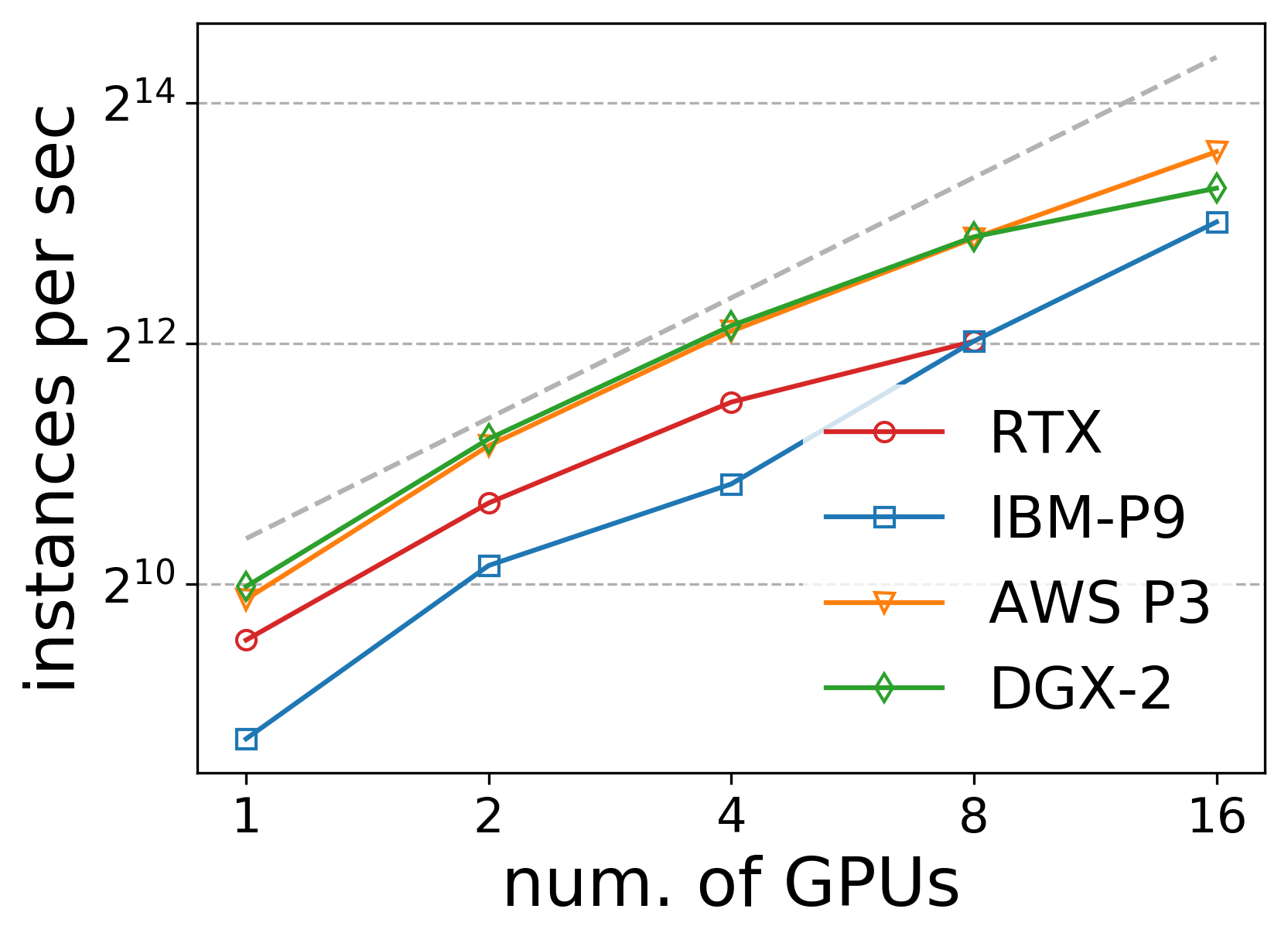}
        \caption{\label{fig:loglog-resnet18}ResNet18}
    \end{subfigure}
    \begin{subfigure}[t]{0.49\columnwidth}
        \includegraphics[width=1\columnwidth]{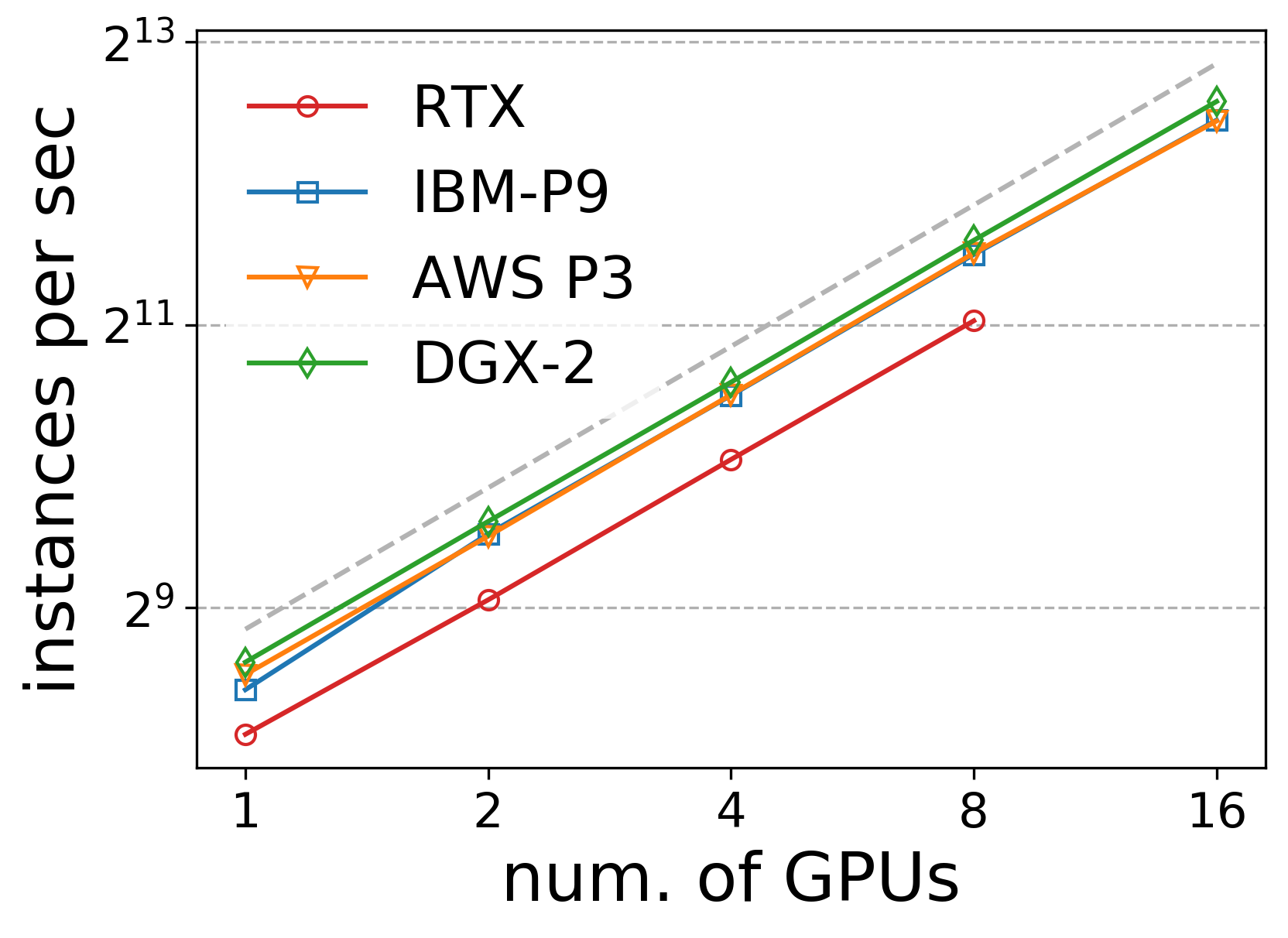}
        \caption{\label{fig:loglog-resnet50}ResNet50}
    \end{subfigure}
    \caption{\label{fig:loglog-SDM}
        Examining Scaling in Log-Log Scale. 
        Gray dashed lines are linear scaling reference lines.
  }
\end{figure*}

\subsubsection{\textbf{Performance Analysis of High-Throughput Learning Models}}\label{subsec:smallDLM}
Here, AlexNet, ResNet18, and ResNet50 are characterized as high-throughput models. All systems except RTX use a 256 batch size per GPU to fully utilize their 32 GB of memory for all models. RTX uses a batch size of 64. Figure~\ref{fig:scale-SDM} illustrates the results. 
%\adf{?????} \ray{
%let's delete this sentence. I tried to use this sentence to summarize, but did
%it in a bad way.}

%Training high-throughput models implies the added cost of data movement through the file system. Because of the large block size and its GPFS file system, the IBM-P9 achieves a throughput ceiling of 4000 instances per second for AlexNet and ResNet18. Its performance on ResNet50 is similar to that of the DGX-2 and AWS P3, which uses local NVMe hard drives and employs a file system with a much smaller block size.  
Training high-throughput models implies frequent data movement through the file system.
For configurations up to 8 GPUs, the performance is lower on IBM-P9.
The reason for that is related to the use of GPFS external filesystem on the
IBM machine, whereas the other system under consideration utilize local storage
for the executation of these small models. 

%\adf{???? CAN USE
%THIS SW FOR ACTUAL BLOCKING? AND HAVE YOU DONE THIS OR IT IS A PHYLOSOPHICAL
%STATEMENT?} \ray{No. I haven't done this in person. This is my guess. Since IBM folks just did 
%performance benchmark using Tensorflow and IBM in-house DDL (distributed deep learning) communication pacakge.
%and the ImageNet data are compressed in TFRecord formats. 
%TFRecord is nothing but Google's ProtoBuf. (An efficient way to compress data). 
%I'm just trying to make some excuses in order to not make IBM look so bad here...}

For ResNet50 (Figure~\ref{fig:scale-resnet50}), all the systems exhibit linear scaling. Because of the ResNet50 model's small size, the slow inter-node Ethernet bandwidth of the AWS P3 does not bottleneck the distributed training throughput performance. 

Because AlexNet uses more than twice the number of parameters of ResNet50, throughput performance is throttled down by the slow Ethernet connection on AWS P3 when two nodes (with a total of 16 GPUs) are in use (Figure~\ref{fig:scale-alexnet}). Even on the DGX-2, AlexNet does not scale linearly to 16 GPUs (shown in Figure~\ref{fig:loglog-alexnet}). When 16 GPUs are in use on the DGX-2, AlexNet spends about 80\% of the active GPU time in communication, whereas ResNet50 spends only about 4\%.

Given its smallest amount of parameters, ResNet18's need for inter-device communication is modest. Even so, as shown in Figure~\ref{fig:loglog-resnet18}, the scaling is not ideal. An interesting observation is that when using 16 GPUs, the AWS P3 performs better than the DGX-2 (Figure~\ref{fig:scale-resnet18}). %The explanation of this complex performance issue follows.

\begin{figure}[ht]
  \begin{subfigure}[t]{0.49\columnwidth}
      \includegraphics[width=1.0\columnwidth]{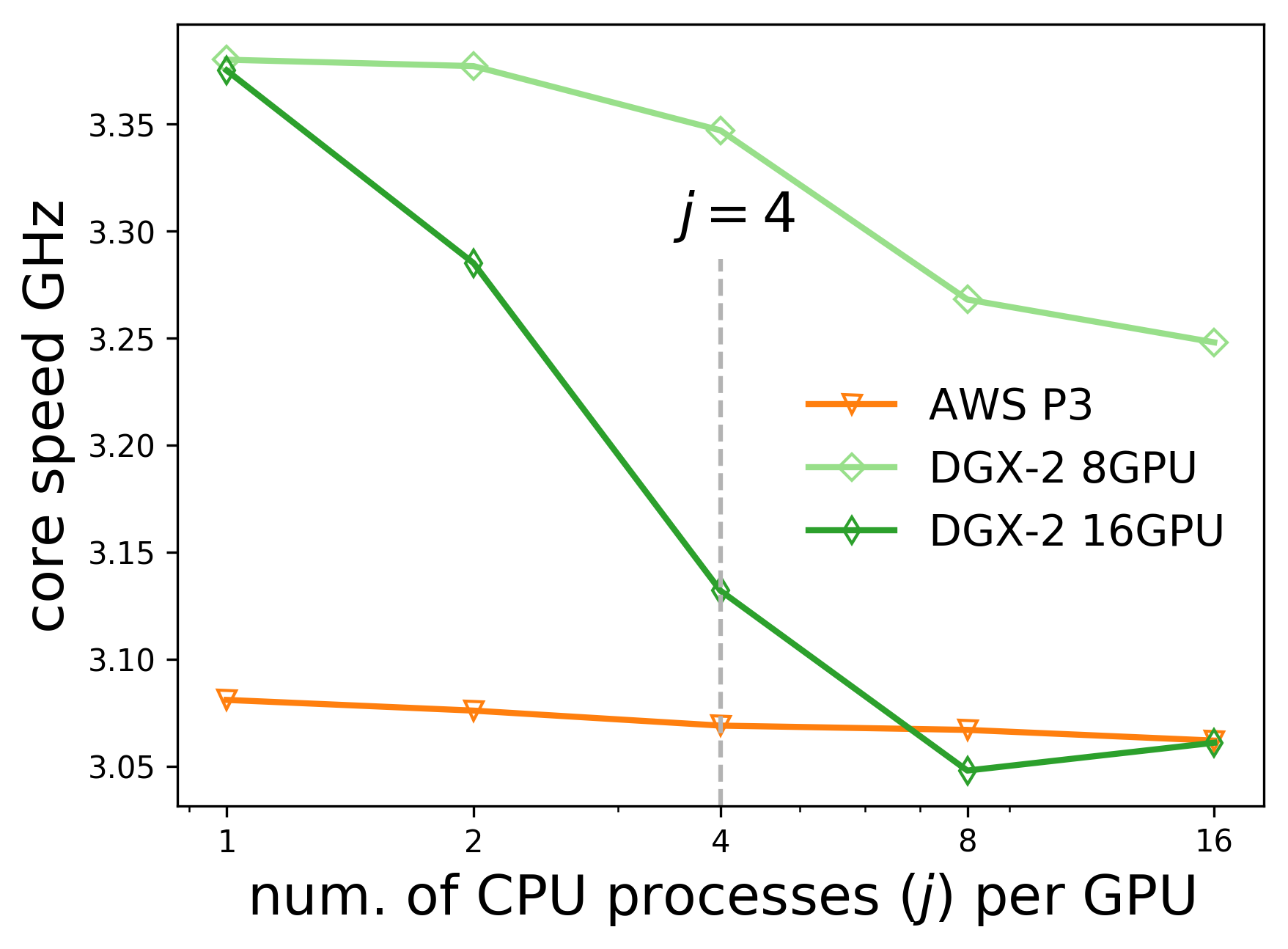}
      \caption{\label{fig:corespeed} CPU Core Speed}
  \end{subfigure}
  \begin{subfigure}[t]{0.49\columnwidth}
      \includegraphics[width=1.0\columnwidth]{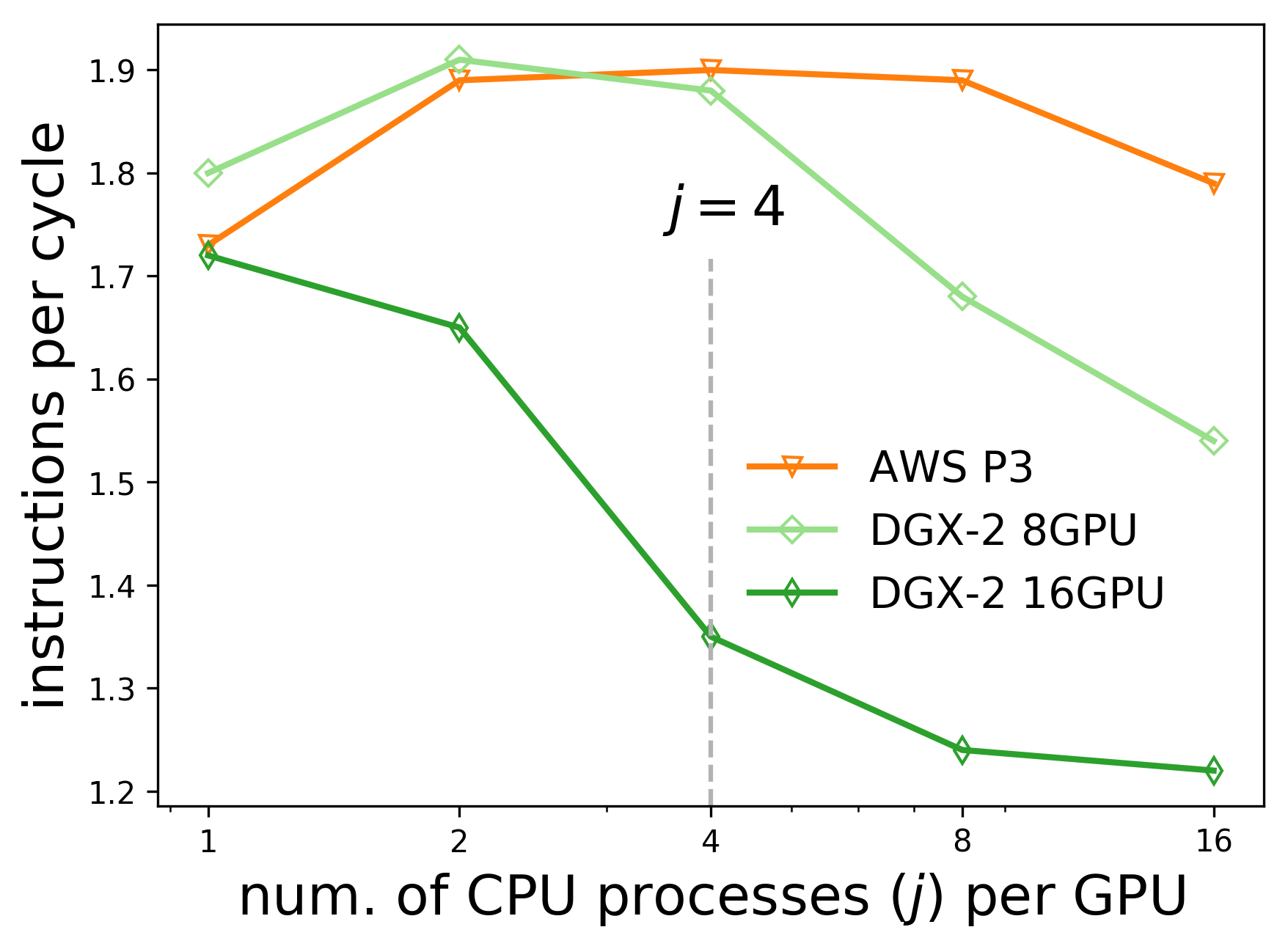}
      \caption{\label{fig:ipc} Instructions per Cycle}
  \end{subfigure}
  \caption{CPU Performance Bottleneck of ResNet18.}
\end{figure}

Recall from Section~\ref{subsec:largeDLM} that in all experiments, each GPU is associated with ($j=$) 4 CPU processes for prefetching data. On the AWS P3, the two CPUs on each node will handle $32$ processes for the eight GPUs. On the DGX-2, the 16 GPUs require $64$ CPU data-fetching 
processes from the two associated CPUs. To explain why the AWS P3 outperforms the DGX-2 in Figure 5b requires determining if the scaling inconsistency stems from a lower core frequency speed and/or cache capacity effects. Figure~\ref{fig:corespeed} shows CPU core speed measurements (enabled given Turbo Boost technology) for both systems while varying $j$ from 1 to 16 on the DGX-2 and AWS P3. For example, if $j=16$ and DGX-2 uses all 16 GPUs, there are 256 CPU processes in total. The light green curve (Figure~\ref{fig:corespeed}) depicts the case when only eight GPUs on the DGX-2 are in use, in which case the DGX-2 has slightly better performance than AWS P3~\ref{fig:scale-resnet18}.
When using $j=1$ CPU process per GPU, the DGX-2's CPU core speed is much higher than that of the AWS P3 because of its superior CPU performance characteristics (see Section~\ref{subsec:hardware}). However, as $j$ increases, the DGX-2's CPU core speed decreases, which is typical for Intel Turbo Boost technology. For $j=4$, the specific case present in the benchmark runs (also shown by the vertical dotted line in Figure~\ref{fig:corespeed}), the DGX-2 maintains a higher CPU core speed than that for the AWS P3. Hence, clock frequency is not the sole explanation for the performance inconsistency. To understand the exact amount of work the CPU does per unit time, Figure~\ref{fig:ipc} shows the metric of instructions per cycle (IPC). The IPC of the DGX-2 using 16 GPUs at $j=4$ is much lower than that of AWS P3: 1.35 versus 1.90, pointing to cache utilization inefficiencies.\footnote{Note: The tested Intel Xeon CPU can reach theoretical maximum of four IPC when instructions are perfectly aligned by manual loop unrolling.} Additional measurements of L1-cache data loading speed and data-translation lookaside buffer (TLB) load misses confirm this hypothesis.%WHY MENTIONING THIS HERE IF WE DON"T SHOW ANY RESULTS? 
The data also reveal that $j=4$ usually is a good choice. Of note, because we use the pinned memory\footnote{Employing pinned memory will prevent the host memory from being swapped out and enable GPU drivers direct access to the host memory.} to improve host-device data transfer, using large $j$ will cause high memory usage on the host.

For RTX versus DGX-2 performance, when one or two GPUs are in use, RTX performance is close to that of the DGX-2 (refer to Table~\ref{tab:rtx-vs-dgx2}). Because of their smaller GPU memory footprints, high-throughput workloads look more suitable on RTX than large models. Just as with the case of performance on large models, RTX's scalability is less than for the DGX-2 (see Table~\ref{tab:rtx-vs-dgx2} and Figure~\ref{fig:loglog-SDM}) due to its slower communication performance. This makes the RTX system most suited for small-scale model development rather than full-scale training workloads.

\subsection{Performance of Mixed-Precision Training}\label{subsec:mixprec}

\begin{figure}[ht]
    \begin{subfigure}[t]{0.49\columnwidth}
        \centering
        \includegraphics[width=1.0\columnwidth]{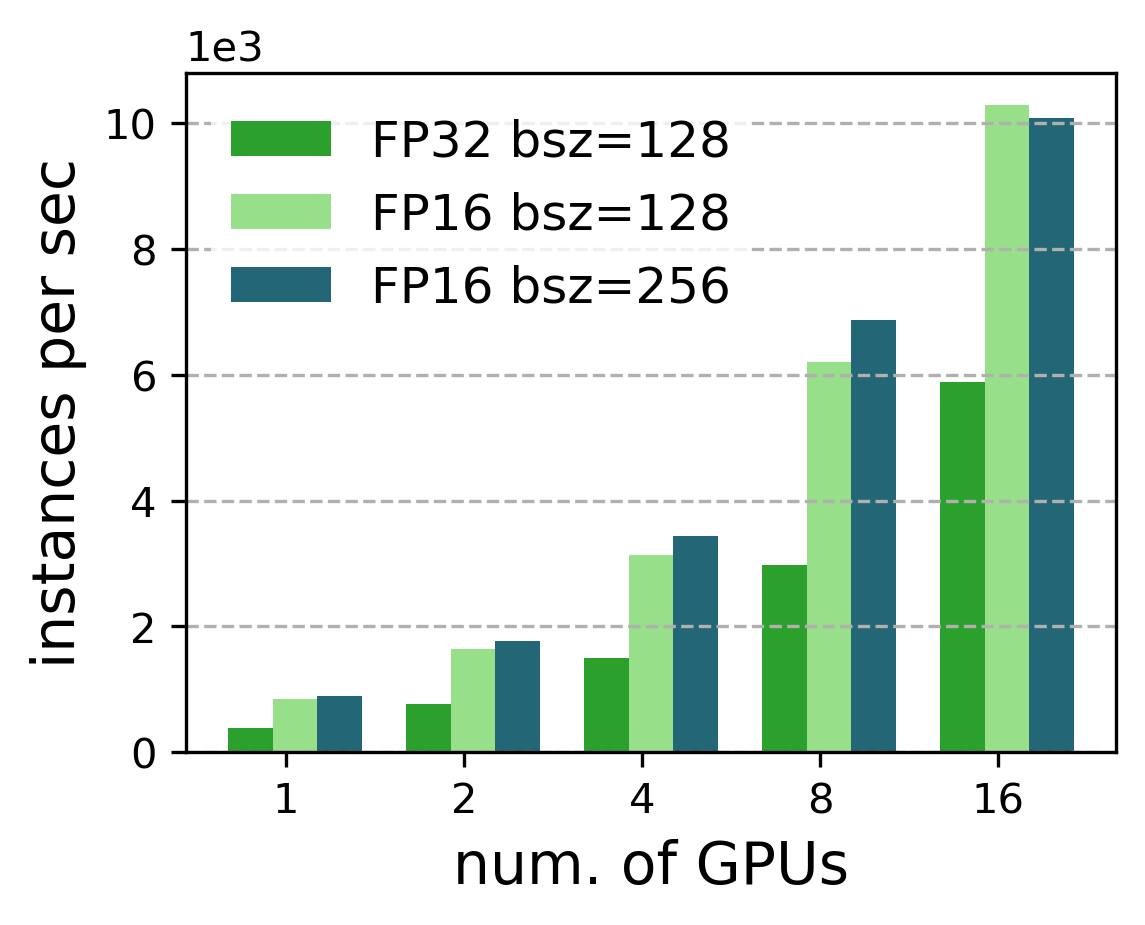}
        \caption{\label{fig:mixedprec:throughput} ResNet50 on DGX-2}
        \vspace*{5mm}
    \end{subfigure}
    \begin{subfigure}[t]{0.49\columnwidth}
        \centering
        \includegraphics[width=1.0\columnwidth]{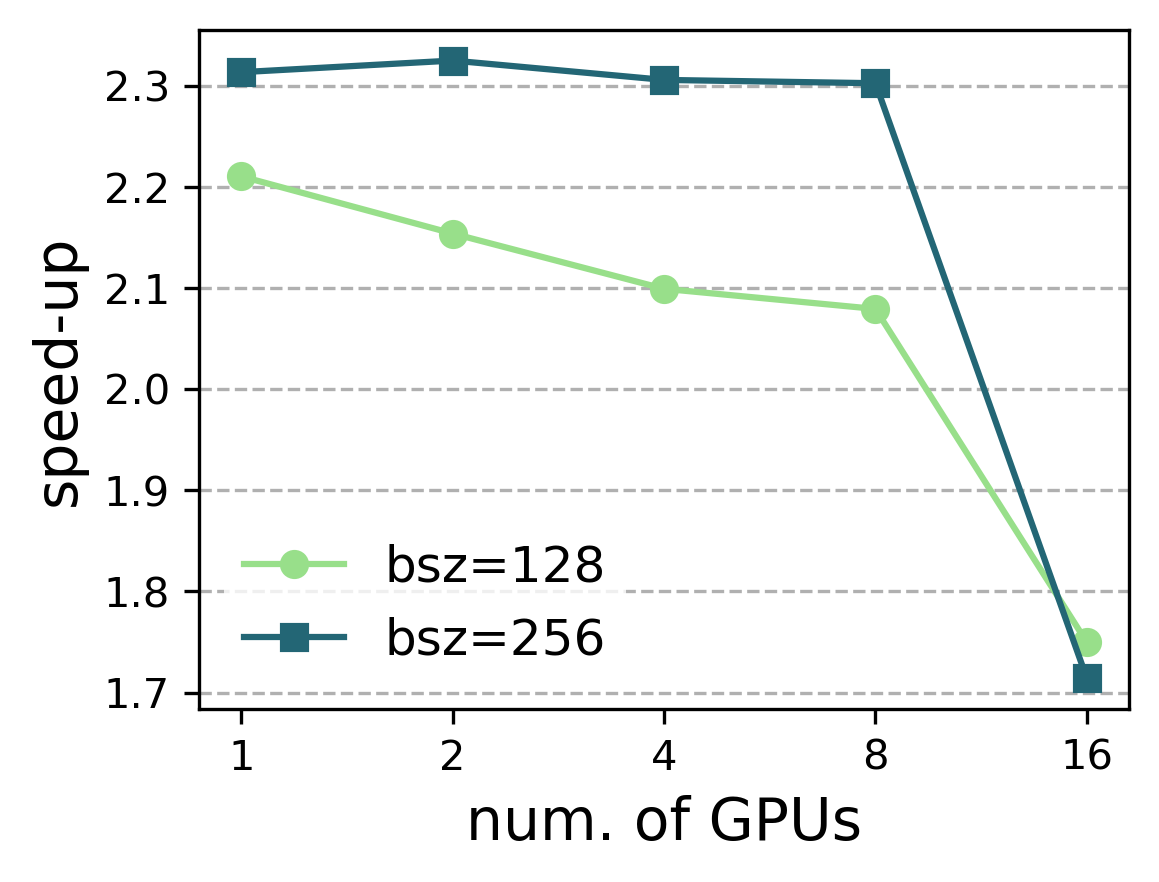}
        \caption{\label{fig:mixedprec:speedup} Speedup using FP16 relative to FP32}
    \end{subfigure}
  \caption{\label{fig:mixedprec} Performance ofResNet50 on DGX-2 for single precision (FP32) and mixed-precision (FP16)}
\end{figure}

Mixed-precision training~\cite{micikevicius2018mixed} retains most if not all
neural network predictive performance, yet offers significant computational
speedup and reduces the memory footprint.  The NVIDIA Turing GPU architecture,
such as V100 and RTX 2080 Ti, provides dedicated hardware acceleration called
``tensor cores''~\cite{nvidia_nvidia_2017-2} for this purpose. The tensor core
provides high-throughput fused multiply-add (FMA) operations for
mixed-precision matrices (inputs in half precision, outputs in either half or
single precision). The other advantage of using mixed-precision is the smaller
memory footprint, therefore less communication overhead for synchronizing the
model replicas.  Figure~\ref{fig:mixedprec:throughput} shows the performance
of ResNet50 on DGX-2 when using mixed-precision (FP16) for batch size
(bsz) 128 and 256, comparing it to the performance when using single-precision (FP32)
for the same model in Figure~\ref{fig:mixedprec:speedup}.
%In all settings except using 16 GPUs, we can
Except for the 16-GPU configuration, we 
achieve more than a factor of 2 performance boost.
%$\times 2$ speedup.  
Moreover, since the memory footprint is smaller for FP16, 
%Since the memory footprint is smaller when using mixed-precision, 
we can accommodate a larger batch size of 256. 
Doubling the batch size halves the synchronization and parameter update time 
for training the same overall amount of data. 
For the 16-GPU configuration, the speedup is only $\times 1.7$. 
This is likely due to the cache effect described in
in Section~\ref{subsec:smallDLM}. 
Note that this performance is very similar to the one reported by NVIDIA~\footnote{https://developer.nvidia.com/deep-learning-performance-training-inference}.

\subsection{Comparing the PyTorch On-node Data Parallel with Distributed Data Parallel}\label{subsec:cdp}

Until now, all of the results herein use the highly optimized distributed data parallel code to achieve the highest system performance possible. By contrast, PyTorch on-node data parallel is an easy-to-use method for enabling computations on multiple GPUs. Code modifications basically are confined to introduction of a directive-like instruction that wraps a non-parallel PyTorch \texttt{Module} with a \texttt{DataParallel} syntax, such as\\ \texttt{model = torch.nn.DataParallel(model).}\footnote{https://pytorch.org/docs/stable/\_modules/torch/nn/parallel/data\_parallel.html.} The communication pattern of on-node data parallel differs from the distributed data parallel. In it, one GPU maintains a master copy of the model parameters. At every iteration, it broadcasts the parameters to the other GPUs in the configuration. At the end of every iteration, the parameters are ``all-reduced'' back to the master GPU, which updates the model parameters. Therefore, for each iteration, two global communications (broadcast and reduce) are issued.  To emulate the common practice of most PyTorch models, we use the default PyTorch data loader for on-node data parallel experiments (\texttt{torch.utils.data.DataLoader}), which supports multi-worker and pinned memory but not asynchronous data loading. PyTorch's on-node data parallel design maximizes its usefulness but targets small parallel GPU configurations, such as those common in workstations.

\begin{figure}[ht]
  \centering
  \includegraphics[width=0.70\columnwidth]{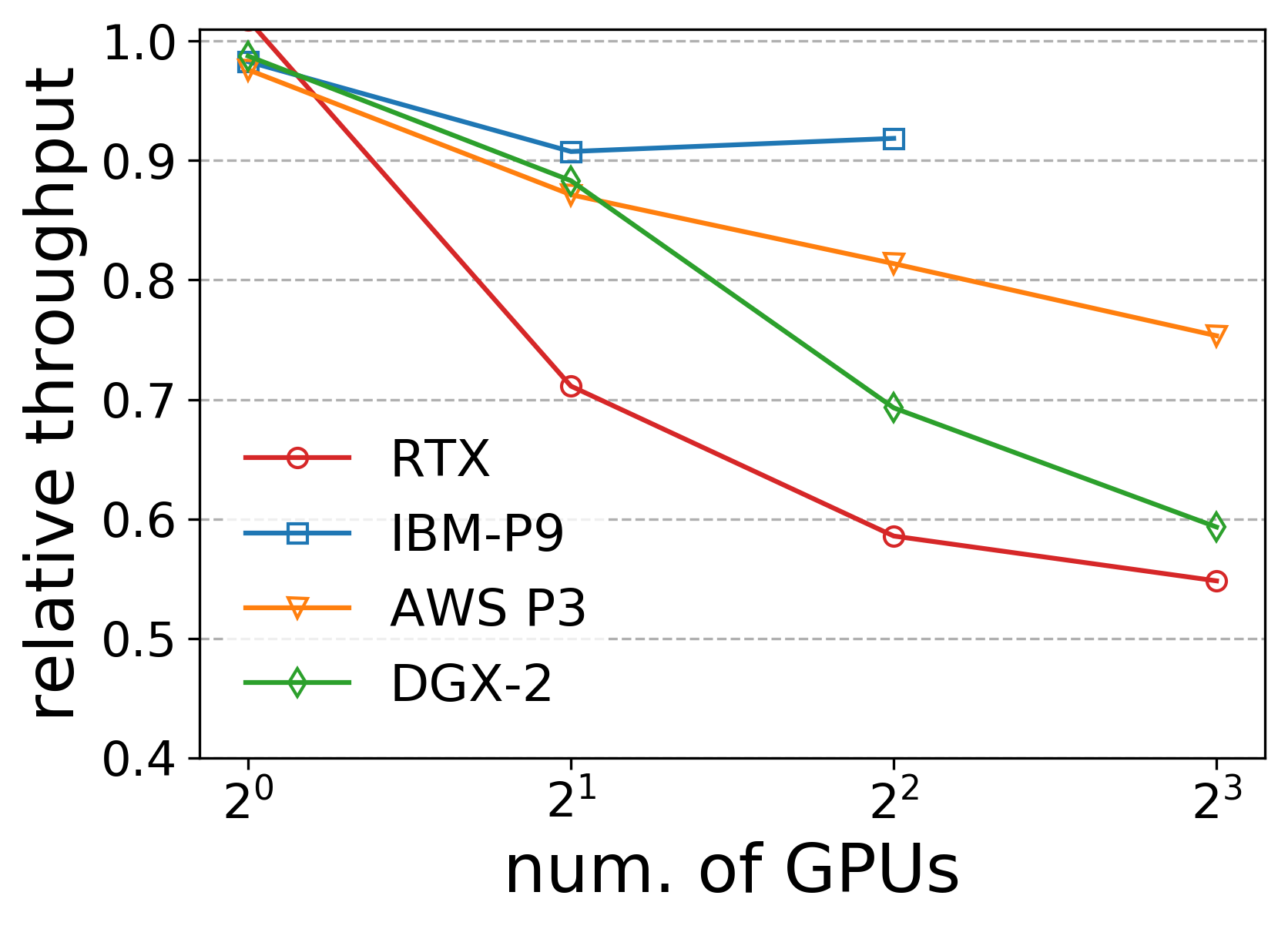}
  \caption{\label{fig:on-node} Relative Throughput Performance of ResNet50 between PyTorch On-node Data Parallel and Distributed Data Parallel.}
\end{figure}

Figure~\ref{fig:on-node} presents the relative performance of models expressed as on-node data parallel compared to the distributed data parallel algorithms for all systems considered. For one GPU, the two data parallel schemes produce similar results. The experiments are done using ResNet50. As more GPUs are utilized, performance decreases when using on-node data parallelism. When two GPUs are in use, DGX-2 and AWS P3 achieve about 90\% of the distributed data parallel performance. Then, it drops rapidly for larger numbers of GPUs. The IBM-P9 can maintain above 90\% upto 4 GPUs.

\section{Conclusion}
In this work we analyzed the performance of several leading-edge systems architected for DL workload performance: DGX-2, AWS P3, and IBM-P9. We also considered a consumer-grade, budget-efficient system: a RTX-2080 Ti server. The inclusion of AWS P3, which essentially is a DGX-1 system, was done to explore performance along the ever-increasing use of cloud computing scenarios for DL workloads. The tested DL models spanned the computer vision and NLP domains, are realistic, and actually are used in real-life DL applications\@. By varying the types of  neural network models and batch sizes per GPU, the systems were probed using different realistic computation and communication scenarios. Some of the specific performance aspects revealed in this work include: 
\begin{itemize}
\item The DGX-2 offered the best 16 GPU collective communication, making it most suited for training large models on 16 GPUs.
\item When training on eight GPUs, the DGX-1, AWS P3, and DGX-2 afforded similar performance. 
\item Because of the limited GPU memory and PCIe bandwidth, when eight GPUs are in use, the RTX-2080 Ti server can reach about 61.46\% of the throughput performance offered by the leading-edge systems considered in this evaluation.
\item The cloud-use scenario is not leading to very large performance degradation when the communication-to-computation ratio of the DL models is low. However, achieving that level of performance requires extensive understanding about the cloud environment to maximize performance by minimizing system contention, ensure geographical closeness of systems, and other idiosyncratic tasks.
\item Scalability of the DL models was investigated up to the sizes of the DGX-2 machine available as a standalone system. Future work will need to consider scaling up to production-size DL models.
\end{itemize}

Practical considerations can be readily extracted from the work documented in this paper, including regarding guidance for procuring systems that maximize performance for a given workload of interest, as well as for considering choice of machines, DL models, and use modes. While as part of this work we implicitly considered cost impacts in system selection, readers are left to weigh such an analysis (and aspects related to it) on their own.

%One aspect left for the imagination and analysis of the reader is that of the system cost, but here too 

% \begin{acks}
\section*{Acknowledgment}
    The authors extend their sincere gratitude to Ethan Hereth (University of
    Tennessee at Chattanooga) for his exhaustive assistance and support related
    to the IBM-P9, as well as Anthony Skjellum (University of Tennessee at
    Chattanooga) for his additional oversight. Too, they thank IBM's Xinghong
    He, Mladen Karcic, and Douglas L. Lehr for facilitating access to internal
    benchmarking resources (four IBM-P9 node configuration) used in this work.
    Thanks also to Brian Barrett (Amazon Web Services) for his assistance
    related to the AWS P3 and to Craig Tierney and Louis Capps (both of NVIDIA)
    and Zhihua Dong (Brookhaven Lab Computational Science Initiative) for DGX-2
    benchmarking support. The authors are grateful for the significant
    assistance received from Charity Plata (Brookhaven Lab) in the editing and
    graphics enhancements of this paper.
    
    This performance analysis was funded as part of the \textit{Exploiting the
    Convergence of Research Challenges in Scientific Discovery And National
Security} program within Brookhaven Lab's Computational Science Initiative with
additional hardware infrastructure support from the Empire State Development
Corporation. Brookhaven National Laboratory is operated and managed for the
U.S. Department of Energy's Office of Science by Brookhaven Science Associates
on behalf of Stony Brook University and Battelle under Contract No.
DE-SC0012704.

% \end{acks}

\bibliographystyle{IEEEtran}
\bibliography{bib/BenchmarkDGX2.bib}
\end{document}